\documentclass[
aps
,groupedaddress
,amsfonts,amsmath,floatfix,eqsecnum
,twocolumn
]{revtex4}

\usepackage[utf8]{inputenc}
\usepackage{epsfig,graphicx}
\usepackage{bm}
\usepackage{amsmath,amsfonts}
\usepackage{enumerate}
\usepackage{ytableau}
\usepackage{cancel}
\usepackage{multirow}
\usepackage{mathptmx}      
\usepackage{latexsym}
\usepackage{amstext}
\usepackage{array}
\usepackage[T1]{fontenc} 

\newcommand{\ignore}[1]{}
\newcommand{\nec}[1]{\eqref{eq:#1}}

\newcommand{\Eq}[1]{Eq.~(\ref{eq:#1})}

\newcommand{\be}{\begin{equation}}
\newcommand{\ee}{\end{equation}}

\def\slashchar#1{\setbox0=\hbox{$#1$}
   \dimen0=\wd0 \setbox1=\hbox{/} \dimen1=\wd1
   \ifdim\dimen0>\dimen1 \rlap{\hbox to \dimen0{\hfil/\hfil}} #1
   \else  \rlap{\hbox to \dimen1{\hfil$#1$\hfil}} / \fi}

\newcommand{\ben}{\begin{enumerate}}
\newcommand{\een}{\end{enumerate}}

\newcommand{\Tr}{{\mathrm{Tr}}}
\newcommand{\tr}{{\mathrm{tr}}}

\newcommand{\cF}{{\mathcal F }}
\newcommand{\cG}{{\mathcal G }}

\newcommand{\cO}{{\mathcal O }}

\newcommand{\cR}{{\mathcal R }}

\newcommand{\hF}{{\hat{F}}}
\newcommand{\hG}{{\hat{G}}}
\newcommand{\hH}{{\hat{H}}}
\newcommand{\hK}{{\hat{K}}}

\newcommand{\bR}{{\mathbf{R}}}

\newcommand{\esp}[1]{\langle #1 \rangle}
\newcommand{\Esp}[1]{\left\langle #1 \right\rangle}

\newcommand{\gf}{\mathrm{gf}}
\newcommand{\gh}{\mathrm{gh}}
\newcommand{\tot}{\mathrm{tot}}
\renewcommand{\div}{\mathrm{div}}
\newcommand{\tg}{\tilde{g}}

\begin{document}

\title{\textsf{
Renormalization of vector fields with mass-like coupling
in curved spacetime
}}

\author{C. Garcia-Recio}
\email{g_recio@ugr.es}

\author{L. L. Salcedo}
\email{salcedo@ugr.es}

\affiliation{Departamento de F\'{\i}sica At\'omica, Molecular y Nuclear and \\
  Instituto Carlos I de F\'{\i}sica Te\'orica y Computacional, \\ Universidad
  de Granada, E-18071 Granada, Spain.
}


\begin{abstract}
Using the method of covariant symbols we compute the divergent part of the
effective action of the Proca field with non-minimal mass term. Specifically a
quantum abelian vector field with a non-derivative coupling to an external
tensor field in curved spacetime in four dimensions is considered. Relatively
explicit expressions are obtained which are manifestly local but non
polynomial in the external fields. Our result is shown to reproduce existing
ones in all particular cases considered. Internal consistency with Weyl
invariance is also verified.
\end{abstract}


\date{\today}


\maketitle
\flushbottom
\setlength{\unitlength}{1mm}

\tableofcontents

\section{\textsf{Introduction}}
\label{sec:1}

Although models involving scalar fields are the most commonly considered in
applications of relativistic gravity and cosmology, e.g. for inflation or
$f(R)$ gravity, vector fields also attract considerable interest
\cite{Turner:1987bw,EspositoFarese:2009aj,Maleknejad:2012fw}. As regards to
introducing a persistent anisotropy after inflation, it has been pointed out
that minimally coupled vector fields would not suffice so non-minimally
coupled models have been considered \cite{Golovnev:2008cf}. In such non
minimal models the vector field can be coupled to a mass-like term
$M^{\mu\nu}(x)$ with a possibly local dependence and a possibly non trivial
tensor structure (see \Eq{2.1}).  Most of these studies are at the classical
level and it is only natural to investigate the effect of quantum
fluctuations. As it turns out, the evaluation of the quantum fluctuations of
vector fields with a non minimal coupling is not entirely straightforward.
For scalar or minimal vector theories, the ultraviolet (UV) divergent part of
the effective action, $\Gamma^\div$, is local (hence a polynomial with respect
to the covariant derivatives) and also a polynomial in the external fields. A
notorious exception to this rule is the metric, due to its coupling to the
kinetic energy term in the action. Nevertheless, locality still requires that
terms involving derivatives of the metric are a polynomial in the curvature
and derivatives of it. At variance with this, for a generalized Proca field
locality is preserved but $\Gamma^\div$ is no longer a polynomial in
$M^{\mu\nu}(x)$.\footnote{Unless, $M^{\mu\nu}(x)= m^2 g^{\mu\nu}$ with
  constant $m$. In this case $\Gamma^\div$ is indeed a polynomial in $m^2$.}

The peculiar behavior is due to the kinetic energy term. As is well-known, in
a direct Lorentz covariant formulation, the kinetic-energy term of an abelian
vector field displays ${\mathrm U}(1)$ gauge invariance. This implies that the
quantum fluctuations are not efficiently quenched for all polarizations,
resulting in a propagator with a singular kernel. A mass term breaks such
gauge invariance and changes the number of propagating degrees of freedom, but
the leading (i.e., most UV divergent) term of the action is still
singular. The mass term introduces a penalty to large amplitude fluctuations
of the vector field, but large wavenumbers are not suppressed for the
longitudinal polarization. When this issue is resolved, removing spurious
degrees of freedom, on finds that $M^{\mu\nu}(x)$ behaves as an additional
metric field.

Early studies of non-minimally coupled vector fields were undertaken in
\cite{Novello:1979ik} at the classical level and in \cite{Davies:1984vm} at
the quantum level. The first explicit attempt to a calculation of
$\Gamma^\div$ for the action in \Eq{2.1} have been addressed in
\cite{Toms:2015fja} using the local momentum approach \cite{Bunch:1979uk}. To
cope with the above mentioned singular-kernel problem, the canonical
quantization scheme of Faddeev and Jackiw \cite{Faddeev:1988qp} was used. The
results in \cite{Toms:2015fja} are partial because only the ultrastatic case
is considered in detail. It is correctly concluded that UV divergences can not
be removed by a local and polynomial (in $M^{\mu\nu}$) counterterm
Lagrangian. A new attempt was taken in \cite{Buchbinder:2017zaa}. Unlike the
canonical quantization approach, manifest relativistic covariance was
preserved by using the St\"uckelberg's method
\cite{Stueckelberg:1900zz,Ruegg:2003ps} to transform the action into one with
exact gauge symmetry. The gauge is then fixed, including the usual
compensating Faddeev-Popov term. The new action contains now a vector field
and a scalar field (plus a ghost field that is completely decoupled from the
other fields). The approach of \cite{Buchbinder:2017zaa} was to diagonalize
the vector-scalar action using a non-local kernel. Unfortunately, as noted in
\cite{Ruf:2018vzq}, the detailed implementation of this step is questionable,
and the resulting divergent part of the effective action turned out to be non
local. A complete and impeccable calculation of $\Gamma^\div$ has been carried
out in \cite{Ruf:2018vzq}, where also the various types of generalized Proca
fields are classified. The calculation there, besides using St\"uckelberg's
method, exploits the Weyl invariance of the action (see Sec. \ref{sec:2}). In
this way the problem is transformed into to one where the external fields are
two metric fields and a heat kernel approach is then applied. The final result
is expressed in terms of the two metric fields (and their corresponding
connection and curvature structures). It is fully local, although not
polynomial in $M^{\mu\nu}(x)$, and several cross-checks are satisfied.

In this work, we also carry out a calculation of the functional
$\Gamma^\div[g_{\mu\nu},M^{\mu\nu}]$ for the action in \Eq{2.1}, starting from
the St\"uckelberg formulation introduced in \cite{Buchbinder:2017zaa}. The
difference between our calculation and that in \cite{Ruf:2018vzq} is that we
use throughout the original metric $g_{\mu\nu}$, with the exception of a term
for which a different metric is clearly superior, and in any case just one
metric is present in each single term of the final result. Another difference
is that, instead of the heat kernel, we use the method of covariant symbols,
which seems quite appropriate for this kind of problems. The method was
introduced in \cite{Pletnev:1998yu} for flat spacetime and extended to curved
spacetime in \cite{Salcedo:2006pv}, and also to finite temperature in
\cite{MoralGamez:2011en}.  It has been applied to fermions
\cite{Salcedo:2000hx,Salcedo:2008bs,GarciaRecio:2009zp,Garcia-Recio:2014ffa}
and to obtain a strict derivative expansion of the heat kernel in curved
spacetime \cite{Salcedo:2007bt}. The method covariant of symbols is related to
the method of symbols (of pseudodifferential operators) as described in
\cite{Nepomechie:1984wt} and \cite{Salcedo:1994qy}, where the shift
$\nabla_\mu\to \nabla_\mu+p_\mu$ is applied and $p_\mu$ represents the
momentum of the particle running in the quantum loop. However, in the method
of covariant symbols results are manifestly covariant (i.e., the covariant
derivative appears only in the form $[\nabla_\mu,~]$) and in this sense it
is closer to the momentum space approach of \cite{Bunch:1979uk}. Indeed,
$p_\mu$ is introduced in such a way that any pseudodifferential operator
constructed out of $\nabla_\mu$ and other multiplicative operators is mapped
to a covariant operator which is multiplicative with respect to $\nabla_\mu$
(although it may contain derivatives with respect to $p_\mu$). This guarantees
that all the expressions are local throughout the calculation and the UV
divergence is controlled by the integration over the loop momentum
$p_\mu$. Moreover the map is an algebra homomorphism, hence the covariant
symbol of any operator is immediately obtained from the covariant symbols of
its building blocks (e.g., $\nabla_\mu$ and $M^{\mu\nu}$). Our final result
avoids the use of a bimetric formulation yet it agrees with previous results
in the literature and in particular it correctly reproduces those in
\cite{Ruf:2018vzq}.

The paper is organized as follows. In Sec. \ref{sec:2} we discuss the
formulation of the problem to make the kernel a regular one at the price of
introducing the St\"uckelberg scalar field. The expansion organizing the
calculation is spelled out, and the Weyl symmetry of the problem is noted.  In
Sec. \ref{sec:3} we present the calculation of the terms which are
elementary. Also there we summarize the method of covariant symbols, which is
already applied in that section for some of the terms. In Sec. \ref{sec:4} the
remaining terms are computed through a systematic use of the method of
covariant symbols. Rather explicit expressions are obtained involving only the
original metric. The number of terms has been minimized using integration by
parts. Some elliptic integrals are left implicit, as more detailed expressions
would not be helpful. Several checks of the result are done in
Sec. \ref{sec:5}, considering particular cases or expansions and the validity
of Weyl invariance of the final expression.  Our conclusions are presented in
Sec. \ref{sec:6}. Further details regarding conventions, proving auxiliary
results, or summarizing covariant symbols properties are presented in the
appendices.

\section{\textsf{Formulation of the problem}}
\label{sec:2}

The goal is to obtain the divergent part of the effective action,
$\Gamma^\div$, of an abelian vector field $A_\mu(x)$ in curved spacetime
coupled to an external tensor field. The divergent part of the effective
action will be extracted using dimensional regularization and Euclidean
signature is used throughout.

The action is given by
\begin{equation}
S = \int d^4x\sqrt{g} \left( \frac{1}{4}\cF^{\mu\nu}\cF_{\mu\nu} +
\frac{1}{2}M^{\mu\nu} A_\mu A_\nu \right)
\label{eq:2.1}
\end{equation}
where
\begin{equation}
\cF_{\mu\nu} = \nabla_\mu A_\nu - \nabla_\nu A_\mu
\,.
\end{equation}
The connection is the Levi-Civita connection for the Riemannian metric
$g_{\mu\nu}$, hence $\cF_{\mu\nu}$ coincides with $\partial_\mu A_\nu -
\partial_\nu A_\mu$. Unless otherwise stated $g_{\mu\nu}$ is used to raise,
lower and contract world indices. $M^{\mu\nu}(x)$ is an abelian symmetric
tensor field which is assumed to be positive definite, so that the Gaussian
functional integration over $A_\mu(x)$ converges for large amplitude
fluctuations.

The kinetic term is gauge invariant, implying that fluctuations with large
wavenumbers are not suppressed for the longitudinal polarization. To cope with
this problem we follow \cite{Buchbinder:2017zaa} and apply St\"uckelberg's
method. A new scalar field $\varphi$ is introduced and the field $B_\mu$ is
defined through the change of variables
\begin{equation}
A_\mu = B_\mu + \frac{1}{m}\nabla_\mu \varphi
\,.
\end{equation}
The mass $m$ is arbitrary and is introduced so that the $\varphi$ has the
standard dimensions. Since it can be reabsorbed in the field and its value has
no effect on the final result (as is readily verified) we set $m=1$ from now
on.  In the new variables the action takes the form
\begin{equation}\begin{split}
S &= \int d^4x \sqrt{g} \,\Big( \frac{1}{4}\cF^{\mu\nu}\cF_{\mu\nu} +
\frac{1}{2}M^{\mu\nu} B_\mu B_\nu 
+ M^{\mu\nu}B_\mu\nabla_\nu\varphi
\\& \quad
+ M^{\mu\nu} \nabla_\mu\varphi \nabla_\nu\varphi
\Big)
\end{split}\end{equation}
and $\cF_{\mu\nu} = \nabla_\mu B_\nu - \nabla_\nu B_\mu$.  The whole action is
now gauge invariant (namely, under $B_\mu\to B_\mu + \partial_\mu\Lambda$,
$\varphi\to\varphi-m\Lambda$) since $A_\mu$ is. The next step is to fix the
gauge. A convenient choice is obtained by adding the term
\begin{equation}
S_\gf = \int d^4x \sqrt{g}\, \frac{1}{2} (\nabla^\mu B_\mu)^2
\end{equation}
as well as the compensating Fadeev-Popov ghost term
\begin{equation}
S_\gh =  \int d^4x \sqrt{g} \, \nabla^\mu \omega^* \nabla_\mu \omega
\end{equation}
where $\omega(x)$ is a scalar complex fermionic field.

The total action $S_\tot = S + S_\gf + S_\gh$
can be expressed as
\begin{equation}
S_\tot = S_\gh + 
\int d^4x \sqrt{g}\, \frac{1}{2} \phi^\dagger \hK \phi
\end{equation}
with 
\begin{equation}
\phi = \begin{pmatrix} B_\mu  \\ \varphi \end{pmatrix}
,
\qquad
\hK = \begin{pmatrix} \hF^{\mu\nu} & \hH^\mu \\ \hH^{\dagger\mu} & \hG \end{pmatrix}
.
\end{equation}
The differential operators $\hF$, $\hH$ and $\hG$ are given by
\begin{equation}\begin{split}
\hF^{\mu\nu} &= - g^{\mu\nu} \square + \cR^{\mu\nu} + M^{\mu\nu},
\qquad
\hG = - \nabla_\mu M^{\mu\nu} \nabla_\nu ,
\\
\hH^\mu &= M^{\mu\nu} \nabla_\nu ,
\qquad \hH^{\dagger\mu} =  -  \nabla_\nu M^{\mu\nu} .
\end{split}\end{equation}
where $\square \equiv \nabla^\mu\nabla_\mu$ and $\cR_{\mu\nu}$ is Ricci's
tensor. This tensor is generated from $\frac{1}{4}\cF_{\mu\nu}^2+\frac{1}{2}
(\nabla^\mu B_\mu)^2$, using integration by parts to give $-\frac{1}{2}B^\mu
\square B_\mu - \frac{1}{2} B_\mu[\nabla^\mu,\nabla^\nu]B_\nu$ up to boundary
terms. The operator $\hF$ acts on the space of vectors, while $\hG$ acts on
the space of scalars.

Functional integration over $B_\mu$, $\varphi$ and $\omega$ provides the
effective action
\begin{equation}
\Gamma = \Gamma_K + \Gamma_\gh
,\qquad
\Gamma_K= \frac{1}{2}\Tr\,\log \hK
,
\quad
 \Gamma_\gh = - \Tr_0\log(-\square).
\end{equation}
The subindex zero in $\Gamma_\gh$ indicates to take the functional trace in
the space of scalars.

In order to compute $\Gamma_K$ we split $\hK$ as
\begin{equation}
\hK = \hK_D + \hK_A,\qquad
\hK_D = \begin{pmatrix} \hF & 0 \\ 0 & \hG \end{pmatrix}
,\quad
\hK_A = \begin{pmatrix} 0 & \hH \\ \hH^\dagger & 0 \end{pmatrix}
.
\end{equation}
This allows to make the expansion
\begin{equation}\begin{split}
\Gamma_K &= \sum_{n=0}^\infty \Gamma_{K,n}
,\qquad
\Gamma_{K,0} = \frac{1}{2}\Tr\log\hK_D
,
\\
\Gamma_{K,n} &= \frac{(-1)^{n+1}}{2n} \Tr((\hK_D^{-1}\hK_A)^n) \quad (n>0)
.
\end{split}
\label{eq:2.12}
\end{equation}
In this expansion all terms with odd $n$ vanish since $\hK_A$ has to appear an
even number of times to have a non null contribution to the trace. In
addition, terms with $n > 4$ are UV convergent, so only
$\Gamma_{K,n}$ for $n=0,2,4$ have a contribution to $\Gamma^\div$:
\begin{equation}
\Gamma_K^\div = \Gamma_{K,0}^\div + \Gamma_{K,2}^\div + \Gamma_{K,4}^\div
.
\end{equation}

The zeroth term can be further expanded as
\begin{equation}
\Gamma_{K,0} = \Gamma_F + \Gamma_G,\qquad
\Gamma_F = 
\frac{1}{2}\Tr_1\log\hF ,
\quad
\Gamma_G =  \frac{1}{2}\Tr_0\log\hG 
.
\label{eq:2.13}
\end{equation}

Before finishing this Section, let us note the {\em Weyl symmetry} present in
the action, namely $S$ is invariant under the local reescaling
\begin{equation}\begin{split}
g_{\mu\nu}(x)  &\to g_{\mu\nu}^\Omega(x) = \Omega^2(x) \, g_{\mu\nu}(x),
\\
M^{\mu\nu}(x) &\to (M^\Omega)^{\mu\nu}(x) = \Omega^{-4}(x) \, M^{\mu\nu}(x)
.
\end{split}\end{equation}
This symmetry can be secured in the final result by using Weyl-invariant
combinations, for instance
\begin{equation}
(g^\Omega_{\mu\nu}, (M^\Omega)^{\mu\nu})  = (\hat{g}_{\mu\nu},\tilde{g}^{\mu\nu})
\quad \text{with} \quad
\Omega = (\det(M^\alpha{}_\beta))^{1/8}.
\end{equation}
This choice corresponds to the prescription $\det \hat{g}_{\mu\nu} = \det
\tilde{g}_{\mu\nu}$, where $\tilde{g}_{\mu\nu}$ stands for the inverse matrix
of $\tilde{g}^{\mu\nu}$. This is the approach adopted in
\cite{Ruf:2018vzq}. Here we take the alternative route of using directly the
original pair of external fields
 $(g_{\mu\nu},M^{\mu\nu})$ and Weyl invariance will provide a check of the
calculation. An exception is taken in the case of $\Gamma_G$ since there the
advantages of using $\tilde{g}_{\mu\nu}$ are overwhelming.

\section{\textsf{Elementary contributions to $\Gamma^\div$}}
\label{sec:3}

\subsection{\textsf{$\Gamma_\gh$}}

The value of $\Tr_0\log(-\square)$ is a standard result
\cite{Vassilevich:2003xt} that can be obtained in many ways. In terms of heat
kernel coefficients a well-known relation is
\begin{equation}
\Tr\log(-\square)\big|_\div = \frac{1}{(4\pi)^2}\frac{1}{\epsilon} 
\int d^4x\sqrt{g}\, \tr(b_2(x))
\label{eq:3.1}
\end{equation}
where
\begin{equation}
d = 4+2\epsilon 
\end{equation}
is the dimension parameter in dimensional regularization.\footnote{All our
  calculations are consistent with the results in \cite{Ruf:2018vzq}, up to an
  overall minus sign. This should indicate that $d = 4-2\epsilon$ is being
  used in that reference.} The explicit form of the second Schwinger-DeWitt
coefficient is (see e.g. \cite{Salcedo:2007bt})
\begin{equation}
b_2 = \frac{1}{12}Z_{\mu\nu}^2 + \frac{1}{72}\bR^2 - \frac{1}{180}
  \cR_{\mu\nu}^2 + \frac{1}{180} R_{\mu\nu\alpha\beta}^2
\,.
\label{eq:3.3}
\end{equation}
(See Appendix \ref{app:a} for definitions of the symbols and conventions used
in this work.) This expression of $b_2$ holds for any tensor space. In the
particular case of the scalar space $Z_{\mu\nu} := [\nabla_\mu,\nabla_\nu]$
vanishes, and $\tr_0(1)=1$, hence
\begin{equation}
\tr_0(b_2) = 
\frac{1}{180}\cG 
+ \frac{1}{60} \cR_{\mu\nu}^2 
+ \frac{1}{120}\bR^2 
\,,
\end{equation}
where, following \cite{Ruf:2018vzq}, we have expressed the result using the
topological Gauss-Bonnet term
\begin{equation}
\cG =  \bR^2 - 4 \cR_{\mu\nu}^2 + R_{\mu\nu\alpha\beta}^2
\,.
\end{equation}
The final result for the ghost contribution is therefore
\begin{equation}
\Gamma^\div_\gh = 
\frac{1}{32\pi^2\epsilon}
\int d^4x\sqrt{g}\,
\left(
-\frac{1}{90}\cG 
- \frac{1}{30} \cR_{\mu\nu}^2 
- \frac{1}{60}\bR^2 
\right)
.
\end{equation}

\subsection{\textsf{$\Gamma_G$}}
\label{sec:3b}

The term $\Gamma_G$ can be identified with the effective action corresponding
to the action
\begin{equation}
S_G = 
\int d^4x\sqrt{g}\,
\frac{1}{2} M^{\mu\nu} \nabla_\mu\varphi \nabla_\nu\varphi
\,.
\end{equation}
To deal with this term one approach is that of \cite{Buchbinder:2017zaa} where
$M^{\mu\nu}$ is directly used as an alternative (contravariant)
metric. However, simpler expressions are obtained by using as new
metric $\tg_{\mu\nu}$ that defined through the condition
\cite{Ruf:2018vzq,deRham:2014wfa}
\begin{equation}
\sqrt{g}\,M^{\mu\nu} = \sqrt{\tg}\,\tg^{\mu\nu},
\end{equation}
hence $\tg_{\mu\nu}$ is the inverse of $\tg^{\mu\nu} =
M^{\mu\nu}/\sqrt{\det(M^\lambda{}_\sigma)}$. In this way $S_G$ takes the
standard form
\begin{equation}
S_G = 
\int d^4x\sqrt{\tg}\,
\frac{1}{2} \tg^{\mu\nu} \tilde\nabla_\mu \varphi \tilde\nabla_\nu\varphi
\,.
\end{equation}
This immediately implies that
\begin{equation}
\Gamma_G = \frac{1}{2} \tilde\Tr_0\log(-\tilde{\square})
\end{equation}
and in turn
\begin{equation}
\Gamma^\div_G = 
\frac{1}{32\pi^2\epsilon}
\int d^4x\sqrt{\tg}\,
\left(
\frac{1}{180}\tilde\cG 
+ \frac{1}{60} \tilde\cR_{\mu\nu}^2 
+ \frac{1}{120}\tilde\bR^2 
\right)
,
\label{eq:3.11}
\end{equation}
and of course, $\tg_{\mu\nu}$ is used everywhere in this expression instead of
$g_{\mu\nu}$.

As noted in \cite{Ruf:2018vzq}, the combination $\sqrt{\tg}\,\tilde{\cG}$ can be
replaced with $\sqrt{g}\,\cG$, since its integral is a topological invariant.
Another observation is that this expression is invariant under a global
reescaling of the metric. Since $\tg_{\mu\nu}$ picks up a factor
$1/m^2$ when the mass parameter $m$ is not set to unity, the invariance
property checks that the value of $m$ is not relevant here.

We emphasize that a single metric (and its derived structures) will be used in
any single contribution to the effective action. Only $\tg_{\mu\nu}$ appears
in $\Gamma_G^\div$ while only $g_{\mu\nu}$ appears in our formulas for all the
remaining terms of $\Gamma^\div$.

\subsection{\textsf{$\Gamma_F$ (part 1)}}
\label{sec:3c}

The expression for $\Gamma_F$ in \nec{2.13} also follows from the second
heat-kernel coefficient for the operator $\hF^{\mu\nu}$ and could be borrowed
directly from the results in the literature, however we will evaluate it here
in order to introduce the technique of covariant symbols to be exploited in
the computation of $\Gamma_{K,2}$ and $\Gamma_{K,4}$.

Let us split  $\hF^{\mu\nu}$ into two terms as
\begin{equation}
\hF^{\mu\nu} = -g^{\mu\nu}\square + Y^{\mu\nu},
\qquad
Y^{\mu\nu} = \cR^{\mu\nu} + M^{\mu\nu}
\end{equation}
and apply an expansion in powers of $Y^{\mu\nu}$
\begin{equation}\begin{split}
\Gamma_F &= \frac{1}{2}\Tr_1\log (-g^{\mu\nu}\square + Y^{\mu\nu})
\\
&= \frac{1}{2}\Tr_1\log (-\square) 
- \sum_{n=1}^\infty \frac{1}{2n}\Tr_1((\square^{-1}Y^\mu{}_\nu)^n)
\\
& := \sum_{n=0}^\infty \Gamma_{F,n}
\,.
\end{split}\end{equation}
The UV divergent part of this series finishes at $n=2$ since terms $n\ge
3$ are already UV convergent:
\begin{equation}
\Gamma_F^\div =  \Gamma_{F,0}^\div + \Gamma_{F,1}^\div + \Gamma_{F,2}^\div
.
\end{equation}

For $\Gamma_{F,0}^\div$ \nec{3.1} applies (with $\Tr_1$ and $\tr_1$) as well as
\nec{3.3}. There the term with $Z_{\mu\nu}$ no longer vanishes, instead (using
\nec{A.1})
\begin{equation}
\tr_1(Z_{\mu\nu}^2) = -R_{\mu\nu\alpha\beta}^2.
\end{equation}
Together with $\tr_1(1)=4$, this yields the result
\begin{equation}
\Gamma_{F,0}^\div = \frac{1}{32\pi^2\epsilon} \int d^4x \sqrt{g}
\left(
-\frac{11}{180}\cG 
- \frac{4}{15} \cR_{\mu\nu}^2 
+ \frac{7}{60}\bR^2 
\right)
.
\end{equation}

In order to compute the remaining terms $\Gamma_{F,1}^\div$ and $\Gamma_{F,2}^\div$
we will apply the method of covariant symbols.

\subsection{\textsf{Aside: Covariant symbols}}
\label{sec:3d}

For an operator $\hat{\cO} = \cO(Y,\nabla_\mu)$ constructed with one or more
multiplicative operators $Y(x)$ and the covariant derivative $\nabla_\mu$
(which may include all kind of connections, gauge or other) its covariant
symbol is defined as
\begin{equation}
\overline{\cO} := 
e^{-\frac{1}{2}\{\nabla_\mu,\partial^\mu\}} e^{-\xi^\alpha p_\alpha} \hat{\cO}
e^{\xi^\beta p_\beta} e^{\frac{1}{2}\{\nabla_\nu,\partial^\nu\}}
\big|_{\xi^\mu=0}
\,.
\end{equation}
Here $\{,\}$ denotes the anticommutator, $\xi^\mu$ are the Riemann coordinates
with origin at the point $x$ and corresponding to the connection in
$\nabla_\mu$, although we will only consider the Levi-Civita connection
here. In addition
\begin{equation}
\partial^\mu := \frac{\partial}{\partial p_\mu}
\end{equation}
and $p_\mu$ is a momentum variable to be used as integration variable. For
convenience, in order to avoid a proliferation of factors $i$, we use a purely
imaginary $p_\mu$, hence $p_\mu=ik_\mu$ and $k_\mu$ (real) is the actual
integration variable (still we use $d^d\!p$ as notation). Of course, the
operator $\hat\cO$ itself is assumed not to depend on $p_\mu$ or
$\partial^\mu$.

The covariant symbols were introduced in \cite{Pletnev:1998yu} for flat
spacetime and extended to curved spacetime in \cite{Salcedo:2006pv}. The
relevant properties of the covariant symbols are (see \cite{Salcedo:2006pv}
for details):

1) $\overline{\cO}$ is a covariant multiplicative operator with respect to
$x$, although contains derivatives with respect to $p_\mu$. In addition
$\overline{\cO^\dagger}= (\overline{\cO})^\dagger$, hence
when $\cO$ is hermitian its covariant symbol is also hermitian.

2) The map $\hat{\cO} \to \overline{\cO}$ is an algebra homomorphism, since it
is defined from a similarity transformation. This implies that
\begin{equation}
\overline{f(\cO_1,\ldots,\cO_n)}  = 
f(\overline{\cO_1},\ldots,\overline{\cO_n})
,
\end{equation}
and in particular $\overline{\cO} = \cO(\overline{Y},\overline{\nabla}_\mu)$.
Note also that $\overline{g}_{\mu\nu} = g_{\mu\nu}$ for the Levi-Civita
connection, hence $\overline{(g^{\mu\nu}A_\nu)} = g^{\mu\nu}\overline{A}_\nu$,
etc.

3) The diagonal matrix elements can be rewritten as
\begin{equation}
\esp{ x| \hat\cO |x} = \frac{1}{\sqrt{g(x)}}\int
\frac{d^d \! p}{(2\pi)^d} \overline{\cO}(x,p)
,
\label{eq:3.20}
\end{equation}
where it is understood that all $\partial^\mu$ at the rightmost position
vanish (and also at the leftmost position, from integration by
parts). Therefore, the relation
\begin{equation}
\Tr (\hat\cO) = \int d^4x \sqrt{g(x)} \, \tr(\esp{ x| \hat\cO |x})
\end{equation}
implies
\begin{equation}
\Tr (\hat\cO) 
= \int
\frac{d^4 x \, d^d\!p}{(2\pi)^d} \tr( \overline{\cO}(x,p) )
.
\label{eq:3.22}
\end{equation}
Eqs. (\ref{eq:3.20}) and (\ref{eq:3.22}) are key relations in the
covariant symbol technique, allowing to compute diagonal matrix elements of
operators or functional traces as those appearing in the effective action.

For short we will introduce the notations
\begin{equation}
\esp{f}_x \equiv \int d^4x\sqrt{g}\, f,
\qquad
\esp{f}_p \equiv \frac{1}{\sqrt{g}}\int \frac{d^d \! p}{(2\pi)^d} f
\label{eq:3.23}
\end{equation}
as well as $\esp{f}_{x,p} \equiv \esp{\esp{f}_p}_x$, so that
\begin{equation}
\esp{x|\hat\cO|x} = \Esp{\overline{\cO}}_p,\qquad
\Tr (\hat\cO) 
= \Esp{  \tr(\overline{\cO}) }_{x,p}
.
\label{eq:3.24}
\end{equation}

The explicit form of the covariant symbols for basic operators has been
obtained in \cite{Salcedo:2006pv} in a covariant derivative expansion up to
two derivatives for a general connection and to four derivatives when the
Levi-Civita connection in the world sector is selected (but still arbitrary
with respect to gauge or internal indices). The following results are useful
\begin{equation}
\overline{Y} = Y - Y_\alpha \partial^\alpha + \frac{1}{2!} Y_{\alpha\beta}
\partial^\alpha \partial^\beta
- \frac{1}{3!} Y_{\alpha\beta\gamma} \partial^\alpha \partial^\beta
\partial^\gamma + \cdots
\label{eq:3.25}
\end{equation}
Here $Y$ is any operator that is multiplicative with respect to $x$, i.e., not
containing ``free'' $\nabla_\mu$ (all derivatives appear in the form
$[\nabla_\mu,~]$) and not containing ``free'' $Z_{\mu_1\ldots\mu_n}$ (see
Appendix \ref{app:a} for notational conventions). $Y$ may have world indices
and we have used the convention of adding new indices to the left to indicate
covariant derivatives. So if e.g. $Y=A_\alpha$, $Y_{\mu\nu}$ would be
$A_{\mu\nu\alpha} \equiv [\nabla_\mu,[\nabla_\nu,A_\alpha]]$. Furthermore
\begin{equation}
\overline{\nabla}_\mu = p_\mu
+ \frac{1}{2}Z_{\mu\alpha} \partial^\alpha + 
\frac{1}{6}\cR_{\mu\alpha} \partial^\alpha +
\frac{1}{6}R_{\mu\alpha\lambda\beta} p^\lambda \partial^\alpha
\partial^\beta
+ \cdots
\end{equation}
\begin{equation}\begin{split}
\overline{\square} &= p^\mu p_\mu
+ \frac{1}{6}\bR + Z_{\lambda\alpha} p^\lambda \partial^\alpha
-\frac{1}{3}\cR_{\lambda\alpha} p^\lambda \partial^\alpha
\\ & \quad
+
\frac{1}{3}R_{\lambda\alpha\sigma\beta} p^\lambda p^\sigma \partial^\alpha
\partial^\beta
+ \cdots
\end{split}
\label{eq:3.27}
\end{equation}
Fuller expressions can be found in Appendix \ref{app:b} and in
\cite{Salcedo:2006pv}.

The expansions just presented can be organized by the number of covariant
derivatives so that, for instance $Z_{\mu\nu}$, $R_{\mu\nu\alpha\beta}$,
$\cR_{\mu\nu}$ and $\bR$ count as second order, $g_{\mu\nu}$ as zeroth order,
etc. Alternatively one can grade a term counting the number $N_p$ of $p_\mu$
minus the number of $\partial^\mu$ in that term. Hence the expansions for
$\overline{Y}$, $\overline{\nabla}_\mu$ and $\overline{\square}$ start at
orders $N_p=0,1,2$ and have been made explicit through orders $N_p=-3,-1$, and
$0$, respectively. With this convention one can write, for instance, 
\begin{equation}
\overline{Y} = (Y)_0 + (Y)_{-1} + (Y)_{-2} +  (Y)_{-3} + O(p^{-4})
\end{equation}
with
\begin{equation}
(Y)_{-n} = \frac{(-1)^n}{n!} Y_{\alpha_1 \ldots \alpha_n}
\partial^{\alpha_1} \cdots \partial^{\alpha_n}
.
\label{eq:3.25b}
\end{equation}
$N_p$ is an additive index related to the degree of UV divergence of a term.

Within the covariant symbols technique there are no free $\nabla_\mu$ as the
covariant symbols are multiplicative, however, there are $Z_{\mu\nu}$ or more
generally $Z_{\mu_1\ldots\mu_n}$. These quantities are multiplicative with
respect to $x$ but act on world indices, hence they do not commute with
$p_\mu$ and $\partial^\mu$, instead
\begin{equation}
[Z_{\mu_1\ldots\mu_n},p_\alpha ] = R_{\mu_1\ldots\mu_n\alpha\lambda} p^\lambda
,\quad
[Z_{\mu_1\ldots\mu_n},\partial_\alpha ] = R_{\mu_1\ldots\mu_n\alpha\lambda} \partial^\lambda
.
\end{equation}

An often convenient tool to deal with the momentum integral in $\esp{f}_p$ is
to introduce a tetrad field $e^\mu_a(x)$ to make a change of variables from
$p_\mu$ to $k_a$:
\begin{equation}
g^{\mu\nu} = \delta_{ab} e^\mu_a e^\nu_b,
\quad
\delta_{ab} = e^a_\mu e^\mu_b
,\quad
\det{(e^a_\mu)} = \sqrt{g}
,\quad
p_\mu = i k_a e^a_\mu 
\,.
\label{eq:3.44}
\end{equation}
In this way, if $f(p,X)$ is an expression tensorially constructed out of
$p_\mu$ and tensors $X(x)$ (the operators $\partial^\mu$ are assumed to be no
longer present),
\begin{equation}
\esp{f(p,X)}_p = 
\frac{1}{\sqrt{g}} \int\frac{d^dp}{(2\pi)^d} f(p,X)
= 
\int \frac{d^dk}{(2\pi)^d} f(k,X,e)
,
\label{eq:3.45}
\end{equation}
where $f(k,X,e)$ is tensorially constructed out of $X$ and $e_\mu^a$, and the
scalars $k_a$, checking that $\esp{f}_p$ is indeed a tensor. Upon integration
over $k_a$ the result does not depend on the concrete choice of vierbein field.

Another related observation (not discussed in \cite{Salcedo:2006pv}) refers to
derivatives of $p_\mu$. In an expression of the type $\esp{f(p,X)}_p$ where
$f$ no longer contains $\partial_\mu$ and is constructed entirely with $p_\mu$
and other world tensors $X$, the derivative
\begin{equation}
\nabla_\mu \esp{f(p,X)}_p
\end{equation}
is obtained by applying $\nabla_\mu$ only to $X$ (the other tensors in
$f$) and not to $p_\mu$. So, for instance
\begin{equation}
\nabla_\mu \Esp{ p_\nu F(p_\alpha p_\beta M^{\alpha\beta}) }_p
=
 \Esp{ p_\nu p_\alpha p_\beta M_\mu{}^{\alpha\beta} F^\prime }_p
.
\end{equation}
This observation is useful if one wants to
apply integration by parts (with respect to $\nabla_\mu$) in an expression of
the type $\esp{f}_{x,p}$.

The statement would seem rather trivial as $p_\mu$ is an integration
variable. On the other hand, since $p_\mu$ does not commute with
$Z_{\mu\nu}=[\nabla_\mu,\nabla_\nu]$, it follows that $p_\mu$ does not commute
with $\nabla_\mu$ either. Nevertheless the statement holds. This follows from 
\nec{3.45}: the covariant derivative acts on $X$ and $e_\mu^a$, but for any
given point one can choose the tetrad field so that $\nabla_\mu e_\nu^a$
vanishes at that point, so the correct result is obtained by applying the
derivative only to the tensor fields $X$, and this holds at all points. A more
elaborated proof is presented in Appendix \ref{app:c}.

\subsection{\textsf{$\Gamma_F$ (part 2)}}
\label{sec:3e}

We are now in a position to compute $\Gamma_{F,1}^\div$ and $\Gamma_{F,2}^\div$
using the method of covariant symbols.

To compute the UV divergent part of 
\begin{equation}
\Gamma_{F,1} = -\frac{1}{2} \Tr_1(\square^{-1} Y^\mu{}_\nu),
\end{equation}
we use \nec{3.24} to transform it into
\begin{equation}
\Gamma_{F,1} = -\frac{1}{2} \Esp{
\tr_1 \left(\overline{\square}{}^{-1} \overline{Y}{}^\mu{}_\nu \right)
}_{x,p}.
\end{equation}
The expansions in \nec{3.25} for $Y^\mu{}_\nu$ and in \nec{3.27} for $\square$
apply:
\begin{equation}\begin{split}
\overline{Y}_{\mu\nu} &= (\overline{Y}_{\mu\nu})_0  +
(\overline{Y}_{\mu\nu})_{-1} + (\overline{Y}_{\mu\nu})_{-2} + O(p^{-3}),
\\
\overline\square &= (\overline\square)_2 +  (\overline\square)_0 +
(\overline\square)_{-1} + O(p^{-2}),
\end{split}\end{equation}
and hence
\begin{equation}
\overline\square{}^{-1} = (\overline\square)_2^{-1} 
- (\overline\square)_2^{-1} (\overline\square)_0 (\overline\square)_2^{-1} 
+ O(p^{-5})
.
\end{equation}
One can then expand the product $\overline{\square}{}^{-1}
\overline{Y}{}^\mu{}_\nu$. Clearly all terms with odd degree $N_p$ vanish
within $\esp{~}_p$ due to parity. Also terms with $N_p < -4$ are UV
convergent. In fact within dimensional regularization only the terms with
$N_p=-4$ can have a non null contribution. That is\footnote{Strictly speaking
  the right-hand side of \nec{3.35} produces $\Gamma(-\epsilon)=-1/\epsilon +
  O(1)$. The UV part is defined as the pole term.}
\begin{equation}
\Gamma_{F,1}^\div = -\frac{1}{2} \Esp{ \tr_1\left(
( \overline{\square}{}^{-1}
\overline{Y}{}^\mu{}_\nu )_{-4}
\right)}_{x,p}
\label{eq:3.35}
\end{equation}
with
\begin{equation}\begin{split}
( \overline{\square}{}^{-1}
\overline{Y}_{\mu\nu} )_{-4}
&=
(\overline\square)_2^{-1} (\overline{Y}_{\mu\nu} )_{-2}
- (\overline\square)_2^{-1} (\overline\square)_0
(\overline\square)_2^{-1}(\overline{Y}_{\mu\nu})_0
\\
&= -N_g \frac{1}{2!}Y_{\alpha\beta\mu\nu}
\partial^\alpha \partial^\beta
-
N_g (\overline\square)_0 \, N_g Y_{\mu\nu}
.
\end{split}\end{equation}
Here we have introduced the quantity
\begin{equation}
N_g := (-g^{\mu\nu}p_\mu p_\nu)^{-1}
\label{eq:3.40}
\end{equation}
which is positive definite. A further simplification occurs because terms of
the type $\esp{X\partial^\mu}_p$, as well as $\esp{\partial^\mu X}_p$ vanish
identically, hence the first term in $( \overline{\square}{}^{-1}
\overline{Y}_{\mu\nu} )_{-4}$ drops off:
\begin{equation}
\Gamma_{F,1}^\div = \frac{1}{2} \Esp{ \tr_1\left(
N_g (\overline\square)_0 \, N_g Y^\mu{}_\nu
\right)}_{x,p}
.
\end{equation}
The form of $(\overline\square)_0$ can be read off from \nec{3.27}, namely,
\begin{equation}
(\overline\square)_0 =  \frac{1}{6}\bR + Z_{\lambda\alpha} p^\lambda \partial^\alpha
-\frac{1}{3}\cR_{\lambda\alpha} p^\lambda \partial^\alpha
+
\frac{1}{3}R_{\lambda\alpha\sigma\beta} p^\lambda p^\sigma \partial^\alpha
\partial^\beta
.
\end{equation}
The method here is to move the $\partial^\mu$ to the right or to the left, to
exploit the properties $0=\esp{X\partial^\mu}_p = \esp{\partial^\mu
  X}_p$. This can be conveniently done using relations of the type
\begin{equation}
[ \partial^\alpha , N_g ] = 2p^\alpha N_g^2
,\qquad
[\partial^\alpha, [ \partial^\beta, N_g ]] = 2 g^{\alpha\beta} N_g^2 + 8 p^\alpha
p^\beta N_g^3
\,,
\end{equation}
as well as standard angular averages of the type
\begin{equation}
\Esp{p_\mu p_\nu N_g^3}_p = -\frac{1}{4} g_{\mu\nu} \Esp{ N_g^2}_p
.
\end{equation}
These manipulations produce
\begin{equation}
\Gamma_{F,1}^\div = \frac{1}{12}  \Esp{ \bR Y^\mu{}_\mu N_g^2 }_{x,p}
.
\end{equation}

The quantity $\esp{N_g^2}_p$ is UV divergent and can be reduced to a standard
flat-space form using a vierbein field, as previously discussed around
\nec{3.44}.  Hence
\begin{equation}\begin{split}
\esp{N_g^2}_p &= 
\frac{1}{\sqrt{g}} \int\frac{d^dp}{(2\pi)^d} \frac{1}{(-p_\mu^2)^2}
= 
\int \frac{d^dk}{(2\pi)^d}\frac{1}{(k^2)^2}
\\ &
=
\frac{\Gamma(-\epsilon)}{(4\pi)^2}
= - \frac{1}{(4\pi)^2\epsilon} + O(1)
.
\label{eq:3.47}
\end{split}\end{equation}
Here we have simplified the calculation anticipating the result for the UV
divergent part. A more rigorous treatment would use a denominator
$(-p_\mu^2+m^2)$ with $m^2>0$, to avoid infrared divergences. The contribution
from $m^2$ goes into the $O(1)$ terms, as it should since the effect of $m^2$
is subleading in the UV region. This justifies the prescription of taking
directly the terms of order $p^{-4}$ to isolate the UV divergent
contributions.  It is worth noticing that $\esp{N_g^2}_p $ correctly goes to
$+\infty$ as $d \to 4$ from $d<4$, or equivalently $\epsilon\to 0^-$. This
checks that the sign of our calculations is the correct one.

Hence
\begin{equation}
\Gamma_{F,1}^\div = \frac{1}{32\pi^2 \epsilon}
\Esp{ - \frac{1}{6} \bR Y^\mu{}_\mu }_x
.
\end{equation}

The remaining term is also readily computed:
\begin{equation}
\Gamma_{F,2} = -\frac{1}{4} \Esp{ \tr_1 \left( (
\overline{\square}^{-1} \overline{Y}{}^\mu{}_\nu )^2
\right)}_{x,p}
,
\end{equation}
and selecting the terms of order $p^{-4}$
\begin{equation}
\Gamma_{F,2}^\div = -\frac{1}{4} \esp{ N_g^2 Y_{\mu\nu}^2 }_{x,p}
= \frac{1}{32\pi^2 \epsilon}
\Esp{ \frac{1}{2} Y_{\mu\nu}^2 }_{\!x}
.
\end{equation}

In summary, for $\Gamma_F$ one obtains
\begin{equation}\begin{split}
\Gamma_F^\div &= \frac{1}{32\pi^2\epsilon} \left\langle
-\frac{11}{180}\cG 
- \frac{4}{15} \cR_{\mu\nu}^2 
+ \frac{7}{60}\bR^2 
\right.
\\ & \quad
\left.
- \frac{1}{6} \bR Y^\mu{}_\mu
+\frac{1}{2} Y_{\mu\nu}^2
\right\rangle_x
,
\end{split}
\end{equation}
and in terms of $M_{\mu\nu}$, using $Y_{\mu\nu} = \cR_{\mu\nu} + M_{\mu\nu}$,
\begin{equation}\begin{split}
\Gamma_F^\div &= \frac{1}{32\pi^2\epsilon} \int d^4x \sqrt{g} \,\Big(
-\frac{11}{180}\cG 
+ \frac{7}{30} \cR_{\mu\nu}^2 
- \frac{1}{20}\bR^2 
\\
& \quad
- \frac{1}{6} \bR M^\mu{}_\mu
+ \cR_{\mu\nu} M^{\mu\nu}
+ \frac{1}{2} M_{\mu\nu}^2
\Big)
.
\end{split}\end{equation}
This result reproduces that in Eq. (26) of \cite{Ruf:2018vzq} and agrees with
previous literature \cite{Barvinsky:1985an,Fradkin:1981iu}. This completes the
calculation of $\Gamma^\div_{K,0}$.

For convenience we also make explicit the combined result of $\Gamma_\gh^\div$
and $\Gamma_F^\div$:
\begin{equation}\begin{split}
\Gamma_{\gh + F}^\div &= \frac{1}{32\pi^2\epsilon} \int d^4x \sqrt{g} \,\Big(
-\frac{13}{180}\cG 
+ \frac{1}{5} \cR_{\mu\nu}^2 
- \frac{1}{15}\bR^2 
\\
& \quad
- \frac{1}{6} \bR M^\mu{}_\mu
+ \cR_{\mu\nu} M^{\mu\nu}
+ \frac{1}{2} M_{\mu\nu}^2
\Big)
.
\label{eq:3.53}
\end{split}\end{equation}

\section{\textsf{Remaining contributions to $\Gamma^\div$}}
\label{sec:4}

As shown in Sec. \ref{sec:2} the divergent part of the effective action can be
split as
\begin{equation}
\Gamma^\div = \Gamma^\div_\gh + \Gamma_F^\div + \Gamma_G^\div + \Gamma_{K,2}^\div
+ \Gamma_{K,4}^\div
.
\end{equation}
The computation of the last two terms will be undertaken here.

The term $\Gamma_{K,2}^\div$ is defined in \Eq{2.12} as
\begin{equation}
\Gamma_{K,2} = -\frac{1}{4} \Tr( (\hK_D^{-1} \hK_A)^2 )
.
\end{equation}
Expanding in terms of the matrices $\hK_D^{-1}$ and $\hK_A$, and exploiting
the cyclic property of the trace, which is justified for the UV divergent
component, this expression can be brought to the form
\begin{equation}
\Gamma_{K,2} = - \frac{1}{2} \Tr( \hG^{-1} \hH^\dagger \hF^{-1} \hH )
,
\end{equation}
where we have chosen to use an operator acting on the space of scalars. The
method of covariant symbols then yields
\begin{equation}
\Gamma_{K,2} = - \frac{1}{2} \Esp{ \tr_0\left(\,
 \overline{G^{-1}} \,\, 
  \overline{H^\dagger}{}^\mu \overline{F^{-1}_{\mu\nu}} \,\,
  \overline{H}{}^\nu 
\, \right) }_{x,p}
,
\end{equation}
and for the UV divergent part
\begin{equation}
\Gamma_{K,2}^\div = - \frac{1}{2} \Esp{ \tr_0\left(\,
 \overline{G^{-1}} \,\, 
  \overline{H^\dagger}{}^\mu \overline{F^{-1}_{\mu\nu}} \,\,
  \overline{H}{}^\nu 
\, \right)_{-4} }_{x,p}
,
\label{eq:4.5}
\end{equation}
where the subindex $-4$ indicates to retain only the terms of order $p^{-4}$.
The method to select those terms is to use the known expansions of the
covariant symbols for the basic blocks $\nabla_\mu$, $\square$, $M^{\mu\nu}$,
etc, to obtain
\begin{equation}\begin{split}
\overline{F}{}^{\mu\nu} &= (\overline{F}{}^{\mu\nu})_2
+
(\overline{F}{}^{\mu\nu})_0 + O(p^{-1}),
\\
\overline{G} &= (\overline{G})_2 + (\overline{G})_1
+ (\overline{G})_0 +  O(p^{-1}),
\\
\overline{H}{}^\mu &= (\overline{H}{}^\mu)_1 + (\overline{H}{}^\mu)_0
+ (\overline{H}{}^\mu)_{-1} + O(p^{-2})
.
\end{split}
\end{equation}
Furthermore
\begin{equation}\begin{split}
(\overline{F}{}^{\mu\nu})_2 &= - p_\alpha p^\alpha g^{\mu\nu}
=  g^{\mu\nu} N_g^{-1},
\\
(\overline{G})_2 &= - p_\mu p_\nu M^{\mu\nu} \equiv N_M^{-1}
.
\end{split}
\end{equation}
The quantity $N_g$ was defined in \nec{3.40} while $N_M$ has been newly
defined here and is also positive.  This gives
\ignore{
\begin{equation}\begin{split}
\overline{F^{-1}_{\mu\nu}} &= g_{\mu\nu} N_g 
- N_g (\overline{F}_{\mu\nu})_0 \, N_g + O(p^{-5})
,
\\
\overline{G^{-1}} &= N_M - N_M (\overline{G})_1 N_M
- N_M (\overline{G})_0 \,  N_M
\\ & \quad
+ N_M (\overline{G})_1 N_M (\overline{G})_1 N_M
+  O(p^{-5})
.
\end{split}
\end{equation}
}
\begin{equation}\begin{split}
\overline{F^{-1}_{\mu\nu}} &= (\overline{F^{-1}_{\mu\nu}})_{-2}
+
(\overline{F^{-1}_{\mu\nu}})_{-4} 
+  O(p^{-5})
,
\\
\overline{G^{-1}} &= (\overline{G^{-1}})_{-2}
+
(\overline{G^{-1}})_{-3}
+
(\overline{G^{-1}})_{-4}
+  O(p^{-5})
,
\end{split}
\end{equation}
with
\begin{equation}\begin{split}
(\overline{F^{-1}_{\mu\nu}})_{-2} &= g_{\mu\nu} N_g
\\
(\overline{F^{-1}_{\mu\nu}})_{-4} &=  
- N_g (\overline{F}_{\mu\nu})_0 \, N_g 
,
\\
(\overline{G^{-1}})_{-2} &= N_M 
\\
(\overline{G^{-1}})_{-3} &= 
- N_M (\overline{G})_1 N_M
\\
(\overline{G^{-1}})_{-4} &= 
- N_M (\overline{G})_0 \,  N_M
+ N_M (\overline{G})_1 N_M (\overline{G})_1 N_M
.
\end{split}
\end{equation}
Substitution of these expressions in \Eq{4.5} selects seven terms of the type
$(\overline{G^{-1}})_l (\overline{H^\dagger})_m (\overline{F^{-1}})_n
(\overline{H})_p $ with $(l,m,n,p)$ taking values $(-4,1,-2,1)$,
$(-3,1,-2,0)$, $(-3,0,-2,1)$, $(-2,-1,-2,1)$, $(-2,0,-2,0)$, $(-2,1,-4,1)$,
and $(-2,1,-2,-1)$. The last term is actually vanishing since
$(\overline{H}{}^\mu)_{-1}$ is of the form $X\partial^\nu$.

For $\Gamma_{K,4}$ one has similarly
\begin{equation}
\Gamma_{K,4} = 
-\frac{1}{4}
 \Tr( (\hG^{-1} \hH^\dagger \hF^{-1} \hH )^2)
.
\end{equation}
In this case there is just one term of $O(p^{-4})$, namely,
\begin{equation}
\Gamma_{K,4}^\div = - \frac{1}{2} \Esp{ \tr_0\left(\,
\left(
(\overline{G^{-1}})_{-2} (\overline{H^\dagger})_1 (\overline{F^{-1}})_{-2}
(\overline{H})_1
\right)^2
\right)
 }_{x,p}
.
\label{eq:4.11}
\end{equation}

The calculation proceeds\footnote{The manipulations have been carried out
  using code in Mathematica written by the authors.} by carrying out the
derivatives $\partial^\mu$, either to the right, or to the left when this
gives a lower number of terms. Next the operators $Z_{\mu_1\ldots\mu_m}$ are
also moved to the right or the left to exploit the properties
$\tr_0(XZ_{\mu_1\ldots\mu_m})=\tr_0(Z_{\mu_1\ldots\mu_m}X)=0$. This produces
an expression involving only $p_\mu$ inside momentum integrals with powers of
$N_g$ and $N_M$, and other tensors constructed with $M^{\mu\nu}$ and its
derivatives and the Riemann tensor and its derivatives.

Integration by parts can be applied both with respect to $x^\mu$ and with
respect to $p_\mu$ in order to reduce the number of terms in the final
expression. We have chosen to remove terms having $M^{\mu\nu}$ with more than
one covariant derivative. Likewise the identities
\begin{equation}
\partial^\mu N_g^n = 2nN_g^{n+1} p^\mu,
\quad
\partial^\mu N_M^n = 2nN_M^{n+1} M^{\mu\nu} p_\nu,
\label{eq:4.12}
\end{equation}
have been applied in order to bring the expression to one involving only a few
independent momentum integrals. Such procedure yields the following
result\footnote{Of course there is some ambiguity in writing the result due to
  integration by parts. The result presented has 32 terms grouped into four
  structures as regards to momentum integrals. Shorter expressions, still with
  four structures, could exist. There are shorter expressions, namely with 26
  terms, but involving a larger number of different momentum integrals.}
\begin{equation}\begin{split} 
\Gamma_{K,2+4}^\div &= 
\frac{1}{32\pi^2\epsilon}
\big\langle
 I^{1, 1} T^{1,1}
+
I^{1, 2}_{ \mu  \nu } T^{1, 2}_{ \mu  \nu }
\\ & \quad
+
I^{1, 3}_{ \mu \nu \alpha  \beta } T^{1, 3}_{ \mu \nu \alpha  \beta }
+ 
I^{3, 2}_{\mu \nu \alpha \beta \rho \sigma } T^{3, 2}_{ \mu\nu\alpha\beta\rho\sigma }
\big\rangle_x
.
\label{eq:4.13}
\end{split}\end{equation}

The tensors $T^{n,m}_{\mu_1\ldots \mu_k}$ take the following form
\begin{widetext}
\begin{equation} \begin{split}
T^{1, 1} &=
- \frac{1}{8}
 M_{ \mu \mu \nu } M_{ \alpha \alpha \nu } 
- \frac{1}{8}
 M_{ \mu \nu \alpha } M_{ \nu \mu \alpha }
+ \frac{1}{8}
 M_{ \mu \nu } M_{ \mu \alpha } M_{ \nu \alpha } 
\\& \quad
+ \frac{1}{12}
 M_{ \mu \nu } M_{ \alpha \beta } R_{\mu \alpha \nu \beta }
- \frac{1}{24}
 M_{ \mu \nu } M_{ \mu \nu } \bR 
\,,
\end{split}\end{equation}

\begin{equation} \begin{split}
T^{1, 2}_{ \mu \nu } &=
- \frac{1}{12}
 M_{ \mu \alpha } M_{ \alpha \nu \beta } M_{ \beta \rho \rho }
 - \frac{2}{3}
 M_{ \mu \alpha } M_{ \alpha \nu \beta } M_{ \rho \rho \beta }
+ \frac{1}{4}
 M_{ \mu \alpha } M_{ \alpha \beta \beta } M_{ \rho \rho \nu }
\\ & \quad
 - \frac{1}{3}
 M_{ \mu \alpha } M_{ \alpha \beta \rho } M_{ \beta \nu \rho } 
- \frac{1}{96}
 M_{ \alpha \alpha } M_{ \beta \mu \nu } M_{ \beta \rho \rho }
 - \frac{1}{24}
 M_{ \alpha \alpha } M_{ \beta \mu \nu } M_{ \rho \rho \beta }
\\ & \quad
- \frac{1}{48}
 M_{ \alpha \alpha } M_{ \beta \mu \rho } M_{ \beta \nu \rho } 
+ \frac{1}{6}
 M_{ \alpha \alpha } M_{ \beta \mu \rho } M_{ \rho \nu \beta } 
+ \frac{1}{3}
 M_{ \alpha \beta } M_{ \alpha \mu \nu } M_{ \rho \rho \beta} 
\\ & \quad
- \frac{1}{4}
 M_{ \alpha \beta } M_{ \alpha \mu \rho } M_{ \beta \nu \rho } 
+ \frac{1}{8}
 M_{ \alpha \beta } M_{ \alpha \rho \rho } M_{ \beta \mu \nu }
\\ & \quad
+ \frac{1}{8}
 M_{ \mu \alpha } M_{ \nu \alpha } M_{ \beta \rho } M_{ \beta \rho }
- \frac{1}{8}
 M_{ \mu \alpha } M_{ \alpha \beta } M_{ \nu \beta } M_{ \rho \rho }
+ \frac{1}{4}
 M_{ \mu \alpha } M_{ \alpha \beta } M_{ \beta \rho } M_{ \nu \rho }
\\ & \quad
 + \frac{1}{12}
 M_{ \mu \alpha } M_{ \nu \alpha } M_{ \beta \rho } \cR_{\beta \rho }
- \frac{1}{12}
 M_{ \mu \alpha } M_{ \alpha \beta } M_{ \nu \beta } \bR 
\,,
\end{split}\end{equation}

\begin{equation} \begin{split} 
 T^{1, 3}_{ \mu \nu \alpha \beta } &=
+ \frac{1}{2}
 M_{ \mu \rho } M_{ \nu \rho } M_{ \sigma \alpha \lambda } M_{\lambda \beta \sigma } 
- 
 M_{ \mu \rho } M_{ \nu \sigma } M_{ \rho \alpha \lambda } M_{ \sigma \beta \lambda } 
- \frac{2}{3}
 M_{ \mu \rho } M_{ \nu \sigma } M_{ \rho \alpha \lambda } M_{ \lambda \beta \sigma }
\\ & \quad
+ \frac{1}{2}
 M_{ \mu \rho } M_{ \nu \sigma } M_{ \rho \lambda \lambda } M_{ \sigma \alpha \beta } 
- \frac{1}{24}
 M_{ \mu \rho } M_{ \sigma \sigma } M_{ \lambda \nu \rho } M_{ \lambda \alpha \beta }
+ \frac{2}{3}
 M_{ \mu \rho } M_{ \sigma \lambda } M_{ \rho \nu \sigma } M_{ \lambda \alpha \beta } 
\\ & \quad
- \frac{1}{6}
 M_{ \mu \rho } M_{ \sigma \lambda } M_{ \sigma \nu \rho } M_{ \lambda \alpha \beta }
+ \frac{1}{24}
 M_{ \rho \rho } M_{ \sigma \lambda } M_{ \sigma \mu \nu } M_{ \lambda \alpha \beta }
- \frac{1}{12}
 M_{ \rho \sigma } M_{ \rho \lambda } M_{ \sigma \mu \nu } M_{ \lambda \alpha \beta }
,
\end{split}\end{equation}

\begin{equation} \begin{split}
T^{3 , 2}_{\mu \nu \alpha \beta \rho \sigma } &=
-\frac{1}{3}
 M_{ \mu \lambda } M_{ \lambda \nu \alpha } M_{ \beta \rho \sigma }
+ \frac{1}{12} 
 M_{ \lambda \lambda } M_{ \mu \nu \alpha } M_{ \beta \rho \sigma }
.
\end{split}\end{equation}
\end{widetext}

On the other hand the integrals $I^{n,m}_{\mu_1\ldots\mu_k}$ are defined from
the relation
\begin{equation}
\Esp{ N_g^n N_M^m \, p_{\mu_1}\cdots p_{\mu_k} }_p
=
\frac{1}{32\pi^2\epsilon}
\,
I^{n,m}_{\mu_1\ldots\mu_k}
.
\label{eq:4.18}
\end{equation}
In our case $k=2(n+m)-4\ge 0$ and the $I^{n,m}_{\mu_1\ldots\mu_k}$ are UV
finite.

The integrals $I^{n,m}_{\mu_1\ldots\mu_k}$ can be represented in several ways
and are subject to relations among them. However, these elliptic integrals do
not admit a simple closed form. A straightforward way to extract the UV finite
factor is by using $4$-spherical coordinates, with radial coordinate $r=
N_g^{-1/2}$. In this case
\begin{equation}
I^{n,m}_{\mu_1\ldots\mu_{2n+2m-4}}
= \frac{(-1)^{2n+2m+1} }{\pi^2} 
\int d^3\Omega_{\hat{k}} \frac{\hat{k}_{\mu_1}
  \cdots \hat{k}_{\mu_{2n+2m-4} } }{ (\hat{k}_\mu\hat{k}_\nu M^{\mu\nu})^m},
\end{equation}
with $\hat{k}_a^2 =1$, ~$\hat{k}_\mu = \hat{k}_a e^a_\mu$ ~and~ $g_{\mu\nu} =
\delta_{ab} e^a_\mu e^b_\nu$. Some simplification is obtained by going to the
local frame in which $M^\mu{}_\nu(x)$ is diagonal. In this case the relevant
integrals become
\begin{equation}
\hat{I}^{n,m}_{a_1 \ldots a_{2n}}
= 
\int d^3\Omega_{\hat{k}} \frac{\hat{k}_{a_1}
  \cdots \hat{k}_{a_{2n} } }{ (\sum_a M_a \hat{k}_a^2 )^m},
\end{equation}
where $M_a$ are the eigenvalues of $M^\mu{}_\nu$. All these integrals follow
from applying derivatives with respect to the $M_a$ to the generating integral
\begin{equation}
\hat{I}(z)
= 
\int d^3\Omega_{\hat{k}} \, (z - \sum_a M_a \hat{k}_a^2 )^{-1}
.
\end{equation}

Another explicit expression, closer to that in \cite{Ruf:2018vzq}, is derived
in Appendix \ref{app:d}, namely,
\begin{equation}\begin{split}
I^{\,n,m}_{\mu_1\ldots\mu_{2n+2m-4}}
&=
\frac{(-2)^{3-n-m}}{\Gamma(n)\Gamma(m)} 
\int_0^\infty dt 
\frac{t^{m-1}}{\sqrt{\det ((M_t)^\mu{}_\nu)}}
\\ &  \quad \times 
[(M_t^{-1})^{n+m-2}]_{\mu_1\ldots\mu_{2n+2m-4}}
\,.
\end{split}
\label{eq:4.19}
\end{equation}
Here we have defined
\begin{equation}
(M_t)^{\mu\nu} = g^{\mu\nu} + t M^{\mu\nu}
,
\end{equation}
$(M_t^{-1})_{\mu\nu}$ denotes the inverse matrix of $(M_t)^{\mu\nu}$ and
$[(M_t^{-1})^n]_{\mu_1\ldots\mu_{2n}}$ stands for the symmetrized product of
$n$ factors $(M_t^{-1})_{\mu\nu}$ (hence a total of $(n-1)!!$ terms). E.g.
\begin{equation}\begin{split}
[M_t^{-1}]_{\mu\nu} &= (M_t)_{\mu\nu}
,
\\ 
[(M_t^{-1})^2]_{\mu\nu\alpha\beta} &= 
(M_t^{-1})_{\mu\nu} (M_t^{-1})_{\alpha\beta} 
+ (M_t^{-1})_{\mu\alpha} (M_t^{-1})_{\nu\beta} 
\\ & \quad +
(M_t^{-1})_{\mu\beta} (M_t^{-1})_{\nu\alpha}
\,. 
\end{split}\end{equation}

\Eq{4.19} assumes $n,m\ge 1$. The cases $m=0$ and $n=0$ can be worked
out separately, or obtained from the same formulas with the replacements,
respectively,
\begin{equation}
\frac{1}{\Gamma(m)} \to t \delta(t),
\qquad
\frac{1}{\Gamma(n)} \to \frac{1}{t}\delta(1/t),
\end{equation}
and in both cases the Dirac deltas have their support at $0^+$. This
prescription yields, for instance, $I^{\,2,0} = -2$, in agreement with
\Eq{3.47}.

\section{\textsf{Cross-checks of the calculation}}
\label{sec:5}

\subsection{\textsf{Terms with zero and four derivatives}}

The effective action can be decomposed as a sum of terms classified by the
number of covariant derivatives. In particular $\Gamma^\div$ can be decomposed
as
\begin{equation}
\Gamma^\div = \Gamma^{\div (4)} + \Gamma^{\div (2)} + \Gamma^{\div (0)}
,
\end{equation}
into terms with $4$, $2$ and $0$ covariant derivatives. Such classification is
intrinsic since it corresponds to the response under dilatations. Therefore
each term $\Gamma^{\div (n)}$ is well-defined and should coincide among
different calculations. In our calculation $\Gamma^{\div (4)}$ only gets
contributions from $\Gamma_\gh$, $\Gamma_G$ and $\Gamma_F$, while
$\Gamma^{\div (2)}$ and $\Gamma^{\div (0)}$ only get contributions from
$\Gamma_F$ and $\Gamma_{K,2+4}$.

For the two simplest cases of zero and four covariant derivatives we have
checked that our results reproduce those in \cite{Ruf:2018vzq}.

Specifically, for four-derivative terms we find
\begin{equation}\begin{split}
\Gamma^{\div(4)} &= \frac{1}{32\pi^2\epsilon} \Big\langle
-\frac{1}{15}\cG 
+ \chi \left(
\frac{1}{60} \tilde\cR_{\mu\nu}^2 
+ \frac{1}{120}\tilde\bR^2 
\right)
\\ & \quad
+ \frac{1}{5} \cR_{\mu\nu}^2 
- \frac{1}{15}\bR^2 
\Big\rangle_x
,
\end{split}\end{equation}
where
\begin{equation}
\chi = \sqrt{\det(M^{\mu\nu})\det(g_{\mu\nu})}
.
\end{equation}
The result in \cite{Ruf:2018vzq} is expressed in terms of
\begin{equation}
\hat{g}_{\mu\nu}=\chi^{1/2}g_{\mu\nu},
\quad
\tilde{g}^{\mu\nu}=\chi^{-1}
M^{\mu\nu}
.
\end{equation}
and can be written as
\begin{equation}\begin{split}
\Gamma^{\div(4)}_{\rm RS} &= \frac{1}{32\pi^2\epsilon} \Big\langle
-\frac{1}{15}\cG 
+ \chi \Big(
\frac{1}{60} \tilde\cR_{\mu\nu}^2 
+ \frac{1}{120}\tilde\bR^2 
\\ & \quad
+ \frac{1}{5} \hat{\cR}_{\mu\nu}^2 
- \frac{1}{15}\hat{\bR}^2 
\Big)
\Big\rangle_x
,
\end{split}\end{equation}
noting that $\det(\hat{g})=\det(\tilde{g})= \chi^2\det(g)$.  The two
calculations coincide because under a Weyl transformation $g_{\mu\nu} \to
\Omega^2 g_{\mu\nu}$, the combination $\cR_{\mu\nu}^2 - \frac{1}{3}\bR^2$
transforms into $\Omega^{-4} (\cR_{\mu\nu}^2 - \frac{1}{3}\bR^2)$ up to a
total derivative.

Regarding the term involving no covariant derivatives, the result in
\cite{Ruf:2018vzq} can be written as
\begin{equation}\begin{split}
\Gamma^{\div(0)}_{\rm RS} &= \frac{1}{32\pi^2\epsilon} \Big\langle
\frac{1}{4} \tr(M^2) + \chi^{1/2} I^{\mu\nu}_{(2,1)} \Big(
-\frac{1}{4} (M^2)_{\mu\nu} 
\\ & \quad
+ \frac{1}{8}M_{\mu\nu}\tr(M) 
\Big)
\Big\rangle_x
.
\end{split}\end{equation}
The identity
\begin{equation}
\hat{g}_{\mu\nu} + u \tilde{g}_{\mu\nu} =
u\chi g_{\mu\alpha} (M_t)^{\alpha\beta} (M^{-1})_{\beta\nu},
\quad
t=\frac{1}{u\chi^{1/2}}
,
\end{equation}
implies the relation
\begin{equation}
I^{\mu\nu}_{(2,1)}
=
-2\chi^{-1/2} g^{\mu\alpha} I^{2,1}_{\alpha\beta} M^{\beta\nu}
,
\end{equation}
and hence
\begin{equation}\begin{split}
\Gamma^{\div(0)}_{\rm RS} &= \frac{1}{32\pi^2\epsilon} \Big\langle
-\frac{1}{8} I^{2,0} \tr(M^2)
+ I^{2,1}_{\mu\nu} \Big(
-\frac{1}{4} (M^3)^{\mu\nu} 
\\ & \quad
+ \frac{1}{8}(M^2)^{\mu\nu}\tr(M) 
\Big)
\Big\rangle_x
.
\end{split}\end{equation}

This is to be compared with our result, which receives contributions from
$\Gamma_F$ and $\Gamma_{K,2+4}$:
\begin{equation}\begin{split}
\Gamma^{\div(0)} &= \frac{1}{32\pi^2\epsilon} \Big\langle
-\frac{1}{4} I^{2,0} \tr(M^2)
+\frac{1}{8} I^{1,1} \tr(M^3)
\\ & \quad
+ I^{1,2}_{\mu\nu} \Big(
  \frac{1}{8} (M^2)^{\mu\nu} \tr(M^2)
\\ & \quad
- \frac{1}{8} (M^3)^{\mu\nu} \tr(M)
+ \frac{1}{4} (M^4)^{\mu\nu}
\Big)
\Big\rangle_x
.
\end{split}\end{equation}
The two expressions coincide, as follows from the integration-by-parts
identity
\begin{equation}
I^{1,2}_{\mu\nu} (M^n)^{\mu\nu}
=
- I ^{2,1}_{\mu\nu} (M^{n-1})^{\mu\nu} - \frac{1}{2} I^{1,1} \tr(M^{n-1})
,
\end{equation}
as well as ~$M^{\mu\nu} I^{n,m}_{\mu\nu\alpha_1\ldots\alpha_k} =
-I^{n,{m-1}}_{\alpha_1,\ldots\alpha_k}$.

\subsection{\textsf{c-number $M^{\mu\nu}$}}

A c-number $M^{\mu\nu}$ refers to the case
\begin{equation}
M^{\mu\nu} = X^2 g^{\mu\nu}
.
\end{equation}
As noted in \cite{Ruf:2018vzq} the corresponding effective action can be
obtained through a Weyl transformation from the case $M^{\mu\nu} =
g^{\mu\nu}$. The divergent part of the latter has been computed in
\cite{Barvinsky:1985an,Buchbinder:2012wb}.

As a check of our results we particularize them to the c-number case. The
general expressions for $\Gamma_F^\div$ and $\Gamma_{K,2+4}^\div$ become
\begin{equation}\begin{split}
\Gamma_F^\div
 &=
\frac{1}{32\pi^2\epsilon} \Big\langle 
-\frac{11}{180}\cG 
+ \frac{7}{30} \cR_{\mu\nu}^2 
- \frac{1}{20}\bR^2 
+ \frac{1}{3} \bR X^2
+ 2 X^4
\Big\rangle_x
\\
\Gamma_{K,2+4}^\div &=
\frac{1}{32\pi^2\epsilon} \Big\langle 
\frac{1}{6}\bR X^2 -\frac{1}{2}X^4 +3 X_\mu^2
\Big\rangle_x
\end{split}\end{equation}

The form of $\Gamma^\div_{\gh}$ is unchanged. On the other hand,
$\Gamma^\div_G$ can be worked out using the relation $\tilde{g}_{\mu\nu}= X^2
g_{\mu\nu}$ which is a Weyl transformation. The expansion of the various
curvatures (and the determinant) of $\tilde{g}_{\mu\nu}$ in terms of those of
$g_{\mu\nu}$ produces the result
\begin{equation}\begin{split}
\Gamma^\div_G &= 
\frac{1}{32\pi^2\epsilon}\Big\langle
\frac{1}{180}\cG 
+ \frac{1}{60} \cR_{\mu\nu}^2 
+ \frac{1}{120}\bR^2 
\Big\rangle_x
+
\delta\Gamma^\div_G
,
\end{split}\end{equation}
with
\begin{equation}\begin{split}
\delta\Gamma^\div_G &=
\frac{1}{32\pi^2\epsilon}\Big\langle
 \frac{1}{15}\left(
2\frac{X_\mu X_\nu}{X^2} - \frac{X_{\mu\nu}}{X}
\right)\cR_{\mu\nu}
\\ & \quad
- \frac{1}{15}\left(
2\frac{X_{\mu\mu}}{X} + \frac{1}{2}\frac{X^2_\mu}{X^2}
\right)\bR
+\frac{1}{15}\frac{X_{\mu\nu}^2}{X^2}
+
\frac{13}{30}\frac{X_{\mu\mu}^2}{X^2}
\\ & \quad
- \frac{4}{15}\frac{X_{\mu\nu} X_\mu X_\nu}{X^3}
+ \frac{1}{15} \frac{X_{\mu\mu} X_\nu^2}{X^3}
+ \frac{3}{15} \frac{(X^2_\mu)^2 }{X^4}
\Big\rangle_x
.
\end{split}\end{equation}
As it turns out this expression can be much simplified through integration by
parts and Bianchi identities, namely,
\begin{equation}\begin{split}
\delta\Gamma^\div_G &=
\frac{1}{32\pi^2\epsilon}\Big\langle
-\frac{1}{6}\frac{X_{\mu\mu}}{X}\bR
+ \frac{1}{2} \frac{X_{\mu\mu}^2}{X^2}
\Big\rangle_x
.
\end{split}\end{equation}

After collecting the various contributions one obtains
\begin{equation}\begin{split}
\Gamma^\div &= 
\frac{1}{32\pi^2\epsilon}\Big\langle
-\frac{1}{15}\cG 
+ \frac{13}{60} \cR_{\mu\nu}^2 
- \frac{7}{120}\bR^2 
\\ & \quad
+ \frac{1}{2}\bR X^2 + \frac{3}{2}X^4 + 3 X_\mu^2
-\frac{1}{6}\frac{X_{\mu\mu}}{X}\bR
+ \frac{1}{2} \frac{X_{\mu\mu}^2}{X^2}
\Big\rangle_x
,
\end{split}\end{equation}
in agreement with the result quoted in \cite{Ruf:2018vzq}. Actually, it would
have been sufficient to verify just the terms with two covariant derivatives,
as it has already been shown the coincidence between the two calculations for
terms with zero or four derivatives for arbitrary configurations
$(g_{\mu\nu},M^{\mu\nu})$.

\subsection{\textsf{Perturbative expansion}}

Here we discuss the perturbative expansion of our result for $\Gamma^\div$, using the form
\begin{equation}
M^{\mu\nu} = m^2 g^{\mu\nu}  + Y^{\mu\nu}
.
\end{equation}
Terms up to second order in powers of $Y^{\mu\nu}$ are displayed. We consider
only those terms with at most two covariant derivatives. These are the most
interesting ones to check the result as the calculation of $\Gamma_{K,2+4}$ is
the most laborious one. Terms with four derivatives have already been shown to
coincide with results in the literature.

Specifically, from $\Gamma_F$ and $\Gamma^\div_{K,2+4}$ we obtain
\begin{widetext}
\begin{equation}
\Gamma^\div_F = \frac{1}{32\pi^2\epsilon}\Big\langle
2m^4 + \frac{1}{3}m^2 \bR 
+ m^2 Y_{\mu \mu } - \frac{1}{6}Y_{\mu \mu } \bR + Y_{\mu \nu } \cR_{\mu \nu }
+ \frac{1}{2} Y_{\mu \nu } Y_{\mu \nu }
+ O(\nabla^4) + O(Y^3)
\Big\rangle_x
,
\end{equation} 
\begin{equation}\begin{split}
\Gamma^\div_{K,2+4} &=\frac{1}{32\pi^2\epsilon}\Big\langle
- \frac{1}{2}m^4 + \frac{1}{6}m^2\bR 
- \frac{1}{4} m^2 Y_{\mu \mu } + \frac{1}{12} Y_{\mu \mu } \bR 
- \frac{1}{6} Y_{\mu \nu } \cR_{\mu \nu }
+ \frac{1}{16} Y_{\mu \mu } Y_{\nu \nu } 
- \frac{3}{8}  Y_{\mu \nu } Y_{\mu \nu } 
\\ & \quad
+ \frac{1}{m^2}\Big(
\frac{1}{12} Y_{\mu \mu \nu } Y_{\nu \alpha \alpha}
+ \frac{1}{48} Y_{\mu \nu \nu } Y_{\mu \alpha \alpha}
- \frac{1}{24} Y_{\mu \nu \alpha} Y_{\mu \nu \alpha}
+ \frac{1}{4} Y_{\mu \nu \alpha} Y_{\nu \mu \alpha}
+ \frac{1}{12} Y_{\mu \mu } Y_{\nu \alpha} \cR_{\nu \alpha} 
- \frac{1}{48} Y_{\mu \mu } Y_{\nu \nu } \bR 
\\ & \quad
+ \frac{1}{24}Y_{\mu \nu } Y_{\mu \nu } \bR 
- \frac{1}{6} Y_{\mu \nu } Y_{\alpha \beta } R_{\mu \alpha \nu \beta }
\Big)
+ O(\nabla^4) + O(Y^3)
\Big\rangle_x
.
\end{split}\end{equation}
\end{widetext}
The total $\Gamma^\div_F + \Gamma^\div_{K,2+4}$ can be shown to coincide with
the result obtained in \cite{Ruf:2018vzq} after using integration by parts
there to remove $Y^{\mu\nu}$ with two covariant derivatives.

\subsection{\textsf{Weyl invariance}}

As noted at the end of Sec. \ref{sec:2} all pairs of external fields in the
orbit $(\Omega^2 g_{\mu\nu}, \Omega^{-4} M^{\mu\nu})$ have the same effective
action, and such invariance must be present in $\Gamma^\div$. Because
$\tilde{g}_{\mu\nu}$ is already Weyl invariant, $\Gamma^\div_G$ is also
invariant. So we consider the remaining terms.

Since Weyl transformations form a group, it is sufficient to consider
the infinitesimal case, namely, $\Omega(x)= 1+\omega(x)$ and
$O(\omega^2)$ is neglected. The infinitesimal variations of the building blocks
are readily obtained:
\begin{equation}\begin{split}
&
\delta g_{\mu\nu} = 2 g_{\mu\nu} \omega,
\qquad
\delta M^{\mu\nu} = -4 M^{\mu\nu} \omega,
\\ &
\delta R = -2 R \omega - 6 \omega_{\mu\mu}
,
\quad
\delta R_{\mu\nu} = -2 \omega_{\mu\nu} - g_{\mu\nu} \omega_{\alpha\alpha},
\\ &
\delta C_{\mu\nu}{}^\alpha{}_\beta = 0 \quad \text{(Weyl tensor)},
\\&
\delta N_g = 2 N_g\omega ,
\quad
\delta N_M = 4 N_M\omega,
\\ &
\delta M_\mu{}^{\alpha\beta} = 
-4 M_\mu{}^{\alpha\beta} \omega
-2 M^{\alpha\beta} \omega_\mu
-M_\mu{}^\beta \omega^\alpha
-M^\alpha{}_\mu \omega^\beta
\\ & \qquad\qquad
+ g^\alpha{}_\mu M^{\sigma\beta} \omega_\sigma
+ g_\mu{}^\beta M^{\alpha\sigma} \omega_\sigma
.
\end{split}\end{equation}

The terms with $\omega$ without derivatives correspond to a global
transformation. The invariance of the full expression in the global case is
easily checked as it is almost trivial from dimensional counting. Hence the
variations contain only $\omega$ with derivatives. From integration by parts
with respect to $x$ they can be brought to the a form proportional to
$\omega_\mu$. This procedure gives
\begin{equation}\begin{split}
\delta \Gamma^\div_\gh &= \frac{1}{32\pi^2\epsilon} \left\langle 
-\frac{1}{3} \bR_\mu \omega_\mu 
\right\rangle_x
,
\\
\delta \Gamma^\div_F &= \frac{1}{32\pi^2\epsilon} \left\langle 
\left(
\frac{1}{3} \bR_\mu + 2 Y_{\nu\nu\mu}
\right)\omega_\mu \right\rangle_x
,
\\
\delta \Gamma^\div_{K,2+4} &= \frac{1}{32\pi^2\epsilon} \left\langle 
\left(
- 2 Y_{\nu\nu\mu} + \cO_\mu
\right)\omega_\mu \right\rangle_x
.
\end{split}\end{equation}
The quantity $\cO_\mu$ involves integrals of the type
$I^{n,m}_{\mu_1\ldots\mu_{2n+2m-4}}$ and can be shown to vanish identically
using integration by parts in momentum space.\footnote{In this case the
  relations (\ref{eq:4.12}) were not sufficient and the relation
  $\partial^\mu\log N_M = 2N_M M^{\mu\nu} p_\nu$ was required.} Hence
$\delta\Gamma^\div=0$ is verified.

\section{\textsf{Summary and conclusions}
\label{sec:6}}

We have carried out a complete calculation of the UV divergent part of the
action in \Eq{2.1} within dimensional regularization. The full result is
\begin{equation}
\Gamma^\div = \Gamma^\div_G + \Gamma^\div_{\gh+F} + \Gamma^\div_{K,2+4}
,
\end{equation}
where $\Gamma^\div_G$ is given in \Eq{3.11}, $\Gamma^\div_{\gh+F}$ in
\Eq{3.53}, and $\Gamma^\div_{K,2+4}$ in \Eq{4.13}. We have made use of the
method of covariant symbols, instead of the heat-kernel, and avoided the use
of expressions involving two metric fields in the same term, with the aim of
obtaining relatively explicit formulas.  Nevertheless the result is involved
and this cannot be avoided in any calculation. Our results are fully
consistent with those in \cite{Ruf:2018vzq}. Some of our terms are more
explicit while those in \cite{Ruf:2018vzq} are more structured (relying on a
compact bimetric setting), hence both calculation can be regarded as
complementary.

It is noteworthy that the technique used here could have been applied also to
the action after making the change (Weyl transformation) from
$(g_{\mu\nu},M^{\mu\nu})$ to $(\hat{g}_{\mu\nu},\tilde{g}^{\mu\nu})$.  The
result would have been precisely the same as the one we have already obtained,
albeit with $(\hat{g}_{\mu\nu},\tilde{g}^{\mu\nu})$ playing the role of
$(g_{\mu\nu},M^{\mu\nu})$. However, proceeding in this way we would have
missed Weyl invariance as a check of the calculation, since any functional of
$(\hat{g}_{\mu\nu},\tilde{g}^{\mu\nu})$ is Weyl invariant by
construction. This also puts the paradox that (our version of the) functional
$\Gamma^\div[\hat{g}_{\mu\nu},\tilde{g}^{\mu\nu}]$ is constrained by Weyl
invariance, even if the latter is automatically fulfilled.  The resolution of
the paradox is that a simpler expression can be achieved in terms of the
transformed fields using that the two metrics have the same volume element,
i.e., $\det(\hat{g})=\det(\tilde{g})$. Namely, rearranging the derivatives of
$\tilde{g}_{\mu\nu}$ to form the corresponding (difference) connection
$\delta\Gamma^\lambda{}_{\mu\nu}$ and exploiting the property
$\delta\Gamma^\lambda{}_{\mu\lambda}=0$ as done in \cite{Ruf:2018vzq}. Instead
of doing this we have chosen to use the original fields
$(g_{\mu\nu},M^{\mu\nu})$.

Because the method of covariant symbols works for any gauge or internal index
connections, there is no problem of principle to extend this kind of
calculations to other cases involving fermions or non abelian vector fields,
in the latter case provided the singular kernel problem is suitably dealt
with.

\acknowledgments
This work has been partially supported by the Spanish MINECO (grant No.
FIS2017-85053-C2-1-P) and by the Junta de Andaluc{\'\i}a (grant No. FQM-225).
This study has been partially financed by the Consejer{\'\i}a de Conocimiento,
Investigaci\'on y Universidad, Junta de Andaluc{\'\i}a and European Regional
Development Fund (ERDF), ref. SOMM17/6105/UGR.

\appendix

\section{\textsf{Conventions}
\label{app:a}}

\subsection{\textsf{Riemann tensor}}
$R^\mu{}_{\nu\alpha\beta}$ denotes the Riemann tensor,
$\cR_{\mu\nu}=R^\lambda{}_{\mu\lambda\nu}$ the Ricci tensor and
$\bR=g^{\mu\nu}\cR_{\mu\nu}$ the scalar curvature. Furthermore our convention
for the Riemann tensor is such that
\begin{equation}
[\nabla_\mu,\nabla_\nu]A^\alpha =  + R_{\mu\nu}{}^\alpha{}_\lambda A^\lambda
.
\label{eq:A.1}
\end{equation}

\subsection{\textsf{Covariant derivatives}}

By default indices are raised, lowered and contracted with $g_{\mu\nu}$. (An
exception occurs for expressions with tilde in \ref{sec:3b} for the
computation of $\Gamma_G$.) The covariant derivative uses the Levi-Civita
connection corresponding to $g_{\mu\nu}$, up to the same exception just noted.
Covariant derivatives are indicated {\em by adding indices to the left}. Hence
for instance $A_{\mu\nu\lambda}$ denotes $\nabla_\mu\nabla_\nu A_\lambda$
(meaning $[\nabla_\mu,[\nabla_\nu, A_\lambda]]$),
$R_{\lambda\mu\nu\alpha\beta} = \nabla_\lambda R_{\mu\nu\alpha\beta}$,
$\cR_{\lambda\mu\nu}= \nabla_\lambda\cR_{\mu\nu}$, $\bR_\lambda
=\nabla_\lambda\bR$, etc.

\subsection{\textsf{The operator $Z_{\mu\nu}$}}

The curvature bundle is defined as
\begin{equation}
Z_{\mu\nu} = [\nabla_\mu,\nabla_\nu].
\end{equation}
It is an antihermitian multiplicative operator with respect to $x$ which acts
on world indices. For instance,
\begin{equation}
[Z_{\mu\nu},A^\alpha{}_\beta] =  R_{\mu\nu}{}^\alpha{}_\lambda A^\lambda{}_\beta - 
R_{\mu\nu}{}^\lambda{}_\beta A^\alpha{}_\lambda
\,.
\label{eq:A3}
\end{equation}
Correspondingly $Z_{\mu\nu}$ commutes with world scalars.
Higher rank tensors are defined recursively as
\begin{equation}
Z_{\alpha\mu_1\ldots\mu_n} = [\nabla_\alpha,Z_{\mu_1\ldots\mu_n}]
-\frac{1}{2}\{\nabla_\lambda,R_ {\mu_1\ldots\mu_n}{}^\lambda{}_\alpha\}
\,.
\label{eq:A4}
\end{equation}
The second term in this expression is an exception to our previous
to-the-left-indices derivative convention. Such extra term is required to make
$Z_{\mu_1\ldots\mu_n}$ a multiplicative operator. These operators fulfill
relations analogous to \nec{A3}, e.g.
\begin{equation}
[Z_{\mu_1\ldots\mu_n},A^\alpha] =  R_{\mu_1\ldots\mu_n}{}^\alpha{}_\lambda A^\lambda
\,.
\end{equation}
This and similar previous equations assume that $A^\alpha$ has no other
indices besides the world index $\alpha$, otherwise new terms appear at the
right-hand side. \Eq{A4} is unchanged.

\subsection{\textsf{Momentum variables}}

For convenience we use $p_\mu=ik_\mu$ where $k_\mu$ are real but $\int
d^dp$ is used to denote $\int d^d k$ as no confusion should arise.

\section{\textsf{Some results for covariant symbols}
\label{app:b}}

Here we quote expressions for $\overline{\nabla}_\mu$, $\overline{\square}$
and $\overline{Z}_{\mu\nu}$ up to and including four covariant
derivatives. The expression for a multiplicative operator $Y$ not acting on
world indices is that in \Eq{3.25b}. The formulas apply for $\nabla_\mu$
having an arbitrary connection in gauge or other internal labels, and the
Levi-Civita connection for world indices.  Of course, $\overline{\square}$
coincides with $g^{\mu\nu} \overline{\nabla}_\mu \,\overline{\nabla}_\nu$, and
$\overline{Z}_{\mu\nu} = [\overline{\nabla}_\mu,\overline{\nabla}_\nu]$.  All
indices are contracted with the metric $g_{\mu\nu}$ and for clarity we put all
world indices as covariant ones, except those in $\partial^\mu$.  The
covariant symbols have been split as $\overline{\cO} =
\sum_n(\overline{\cO})_n$ where the subindex $n$ in $(\overline{\cO})_n$
indicates the value of $N_p$ of the component, i.e., the number of $p_\mu$
minus the number of $\partial^\mu$. The components can be equally well be
classified by the number of covariant derivatives they contain.
\begin{widetext}
\begin{equation}\begin{split}
(\overline{\nabla}_\mu)_1 &=
p_{\mu }
,
\\
(\overline{\nabla}_\mu)_0 &= 0
\\
(\overline{\nabla}_\mu)_{-1} &=
- \frac{1}{4} \{ Z_{\nu \mu },\partial^{\nu } \}
+ \frac{1}{12} \{ [ Z_{\nu \mu } , p_{\alpha } ] , \partial^{\nu } \partial^{\alpha }\}
,
\\ 
(\overline{\nabla}_\mu)_{-2} &=
+ \frac{1}{6} \{ Z_{\nu  \alpha \mu} , \partial^{\nu } \partial^{\alpha } \} 
- \frac{1}{24} \{ [ Z_{\nu \alpha \mu } , p_{\beta }], \partial^{\nu } \partial^{\alpha } \partial^{\beta } \} 
,
\\ 
(\overline{\nabla}_\mu)_{-3} &=
- \frac{1}{16} \{ Z_{\nu \alpha \beta \mu },\partial^{\nu } \partial^{\alpha } \partial^{\beta } \}
+ \frac{1}{80} \{ [ Z_{\nu \alpha \beta \mu } , p_{\rho } ] ,
 \partial^{\nu } \partial^{\alpha } \partial^{\beta } \partial^{\rho} \}
+ \frac{1}{48} \{ Z_{\nu \alpha }, [ Z_{ \beta \mu } , \partial^{\alpha } ] 
\partial^{\nu } \partial^{\beta } \} 
\\&\quad
- \frac{7}{720} \{ [ Z_{\nu \alpha } , p_{\beta }] , [ Z_{\rho \mu }, \partial^{\alpha}]
\partial^{\nu } \partial^{\beta } \partial^{\rho } \}
.
\end{split}\end{equation}
\begin{equation}\begin{split}
(\overline{\square})_2 &=
p_{\mu } p_{\mu }
,
\\
(\overline{\square})_1 &= 0
,
\\
(\overline{\square})_0 &= 
+ \frac{1}{2} \{ Z_{\mu \nu }, p_{\mu } \partial^{\nu } \} 
- \frac{1}{3} [ [ Z_{\mu \nu } , p_{\mu } ] , \partial^{\nu } ] 
- \frac{1}{6} \{ [Z_{\mu \nu }, p_{\alpha} ] p_{\mu }, 
\partial^{\nu } \partial^{\alpha } \} 
,
\\ 
(\overline{\square})_{-1} &= 
+ \frac{1}{6} \{ Z_{\mu \nu \alpha }, 
\{ p_{\alpha },\partial^{\mu } \partial^{\nu } \} \} 
+ \frac{2}{3} [ Z_{\mu \mu \nu } , \partial^{\nu } ] 
- \frac{1}{12} \{ [ Z_{\mu \alpha \nu } , p_{\beta }] p_{\nu },
\partial^{\mu } \partial^{\alpha } \partial^{\beta } \} 
,
\\ 
(\overline{\square})_{-2} &= 
- \frac{1}{16} \{ Z_{\mu \nu \alpha \beta }, 
\{ p_{\beta },\partial^{\mu } \partial^{\nu } \partial^{\alpha } \} \} 
+ \frac{1}{40} \{ [ Z_{\mu \nu \alpha \beta } , p_{\rho }] p_{\beta },
 \partial^{\mu }\partial^{\nu } \partial^{\alpha } \partial^{\rho } \}
- \frac{1}{16} \{ Z_{\mu \nu }, \{ [Z_{\mu \alpha } , p_{\beta }] ,
\partial^{\nu} \partial^{\alpha } \partial^{\beta } \} \}
\\&\quad
+ \frac{1}{8} \{ Z_{\mu \nu } Z_{\mu \alpha } , \partial^{\nu } \partial^{\alpha } \}
+ \frac{1}{30} \{ [ Z_{\mu \nu } , p_{\alpha }] [ Z_{\mu \beta }, p_{\rho }] ,
\partial^{\nu } \partial^{\alpha } \partial^{\beta } \partial^{\rho } \} 
+ \frac{1}{60} [ Z_{\mu \nu }, \partial^{\nu }] [Z_{\mu \alpha }, \partial^{\alpha } ]
\\&\quad
+ \frac{2}{45} [Z_{\mu \nu }, \partial^{\alpha } ] [Z_{\mu \alpha } , \partial^{\nu } ]
+ \frac{2}{45} [Z_{\mu \nu }, \partial^{\alpha }] [Z_{\mu  \nu } , \partial^{\alpha } ]
+ \frac{1}{3} [Z_{\mu \alpha \nu \alpha } , \partial^{\nu } ] \partial^{\mu }
- \frac{1}{60} [Z_{\mu \nu \mu \alpha } , \partial^{\alpha } ] \partial^{\nu }
+ \frac{1}{40} [Z_{\mu \mu \nu \alpha } , \partial^{\alpha } ] \partial^{\nu }
.
\end{split}\end{equation}
\begin{equation}\begin{split}
(\overline{Z}_{\mu\nu})_0 &= Z_{\mu\nu}
,
\\
(\overline{Z}_{\mu\nu})_{-1} &=  
- \frac{1}{2} \{ Z_{\alpha\mu\nu}, \partial^\alpha \}
,
\\
(\overline{Z}_{\mu\nu})_{-2} &= 
+ \frac{1}{4} \{ Z_{\alpha\beta\mu\nu}, \partial^\alpha \partial^\beta \}
.
\end{split}\end{equation}
\end{widetext}
These expressions are written so that hermiticity is manifest (namely,
$p_\mu$, $\nabla_\mu$, and $Z_{\mu_1\ldots\mu_k}$ are antihermitian while
$\partial^\mu$, and $\square$ are hermitian). Expanded expressions with
symbols $R,Z,p,\partial$ ordered from left to right can be found in
\cite{Salcedo:2006pv}.

\section{\textsf{Derivatives of momentum integrated expressions}
\label{app:c}}

Here we present an alternative proof of the statement noted at the end of
Sec.\ref{sec:3d}, namely, if $f(p,X)$ is tensorially constructed out of $p_\mu$
and tensors $X$ (and $f$ no longer contains free $\partial_\mu$), the
covariant derivative of $\esp{f(p,X)}_p$ follows from applying the derivative
only to the tensors $X$ and not to $p_\mu$. The proof relies on the choice of
the Levi-Civita connection in the covariant derivative, corresponding to the
metric $g_{\mu\nu}$ which also appears in the definition of $\esp{f}_p$
through the factor $1/\sqrt{g}$ in \nec{3.23}.

Clearly it is sufficient to prove the statement just for the case when the
integrand $f(p,X)$ is a scalar. Otherwise, say the integrand is of the form
$f_{\mu\nu}(p,X)$ and tensorially constructed out of $p_\mu$ and $X$. Then one
can construct a scalar $h = C^{\mu\nu}(x) f_{\mu\nu}$, with a generic tensor
$C^{\mu\nu}$, and it is clear that the statement would hold for
$\esp{f_{\mu\nu}}_p$ if and only if it does for $\esp{h}_p$.

For simplicity we consider just the following case
\begin{equation}
I(x) = \frac{1}{\sqrt{g(x)}} 
\int \frac{d^d p}{(2\pi)^d} f(p_\alpha p_\beta B^{\alpha\beta}(x))
\end{equation}
as it is sufficiently general to illustrate the arguments involved. A (first
order) infinitesimal shift $x^\mu \to x^\mu + \epsilon^\mu$, will produce a
change $I \to I+\epsilon^\mu \nabla_\mu I$. Similarly in the integrand,
\begin{equation}\begin{split}
B^{\alpha\beta}(x+\epsilon) &= 
B^{\alpha\beta} + \epsilon^\mu \partial_\mu B^{\alpha\beta}
\\
&=
B^{\alpha\beta}+ \epsilon^\mu ( \nabla_\mu B^{\alpha\beta}
-\Gamma^\alpha_{\mu\lambda} B^{\lambda\beta}
-\Gamma^\beta_{\mu\lambda} B^{\alpha\lambda} )
\,.
\end{split}\end{equation}
Contraction with $p_\alpha p_\beta$ then gives
\begin{equation}
p_\alpha p_\beta B^{\alpha\beta}(x+\epsilon)
=
 p^\prime_\alpha p^\prime_\beta (B^{\alpha\beta} + \epsilon^\mu \nabla_\mu
 B^{\alpha\beta} )
\end{equation}
with
\begin{equation}
p^\prime_\nu \equiv p_\nu -\epsilon^\mu \Gamma^\lambda_{\mu\nu} p_\lambda
\end{equation}
where terms $O(\epsilon^2)$ are neglected everywhere. Changing the integration
variables from $p_\mu \to p^\prime_\mu$ gives a Jacobian
\begin{equation}
d^d \!p = d^d \!p^\prime (1+\epsilon^\mu \Gamma^\lambda_{\mu\lambda} ) 
= d^d \!p^\prime
 (1+\epsilon^\mu \partial_\mu\log\sqrt{g}) 
.
\end{equation}
This factor exactly cancels with that produced by the shift $x^\mu \to
x^\mu+\epsilon^\mu$ in $1/\sqrt{g(x)}$. In summary,
\begin{equation}
\nabla_\mu \esp{p_\alpha p_\beta B^{\alpha\beta}}_p = \esp{p_\alpha p_\beta \nabla_\mu B^{\alpha\beta}}_p
\end{equation}
as advertised.

\section{\textsf{Momentum integrals}
\label{app:d}}

Let us justify \Eq{4.19}. The momentum integrals are
\begin{equation}
\tilde{I}^{\,n,m}_{\mu_1\ldots\mu_k} = 
\frac{1}{\sqrt{g}} \int \frac{d^dp}{(2\pi)^d}
N_g^n N_M^m \, p_{\mu_1}\cdots p_{\mu_k}
\end{equation}
with $k=2n+2m-4$. Using a Schwinger representation for the propagators
\begin{equation}
\begin{split}
\tilde{I}^{\,n,m}_{\mu_1\ldots\mu_k} &= 
\frac{1}{\sqrt{g}} \int_0^\infty du \int_0^\infty dv 
\frac{u^{n-1}}{\Gamma(n)} \frac{v^{m-1}}{\Gamma(m)}
\\ & \quad \times
\int \frac{d^dp}{(2\pi)^d}
e^{-u(N_g^{-1}+m_0^2)} e^{-v(N_M^{-1}+m_0^2)}
p_{\mu_1}\cdots p_{\mu_k}
.
\end{split}
\end{equation}
To avoid trivial infrared divergences we have introduced a mass $m_0>0$. This
does not modify the UV divergence. Next we rescale $p_\mu\to p_\mu/\sqrt{u}$,
and make a change of variables from $v$ to $t=v/u$. This gives
\begin{equation}
\begin{split}
\tilde{I}^{\,n,m}_{\mu_1\ldots\mu_k} &= 
\frac{1}{\sqrt{g}} 
\int_0^\infty dt \frac{t^{m-1}}{\Gamma(n)\Gamma(m)}
\int_0^\infty du 
\frac{e^{-u(1+t)m_0^2} }{u^{1+\epsilon}}
\\ & \quad \times
\int \frac{d^dp}{(2\pi)^d}
e^{-N_g^{-1} -tN_M^{-1}}
p_{\mu_1}\cdots p_{\mu_k}
.
\end{split}
\end{equation}
Upon integration over $u$ to yield the UV pole and setting $d\to 4$ in
the remaining terms:
\begin{equation}\begin{split}
\tilde{I}^{\,n,m}_{\mu_1\ldots\mu_k} &= 
\frac{\Gamma(-\epsilon)}{\sqrt{g}} 
\int_0^\infty dt 
\frac{t^{m-1}}
{\Gamma(n)\Gamma(m)}
\\ & \quad \times
\int \frac{d^4p}{(2\pi)^4}
e^{-N_g^{-1} -tN_M^{-1}}
p_{\mu_1}\cdots p_{\mu_k}
.
\end{split}\end{equation}
The momentum integral is now standard after Wick's theorem, with exponential
factor $\exp(p_\mu p_\nu (g^{\mu\nu} + t M^{\mu\nu}))$. The factor
$1/\sqrt{\det(g_{\mu\nu})}$ combines with $1/\sqrt{\det((M_t)^{\mu\nu})}$ to
yield the result quoted in \Eq{4.19}. Alternatively, one can use a
tetrad to integrate over $k_a$ instead of $-ip_\mu$, with the same effect.

\end{document}